\documentclass[aps,twocolumn,groupedaddress,floatfix,preprintnumbers,nofootinbib,eqsecnum]{revtex4}
\pdfoutput=1  

\usepackage{amsmath, float, graphicx,placeins,amssymb,multirow}

\def\IR{\relax{\rm I\kern-.18em R}}
 \font\cmss=cmss10 \font\cmsss=cmss10 at 7pt
\def\IQ{\relax{\rm I\kern-.18em Q}}
\def\IZ{\relax\ifmmode\mathchoice
 {\hbox{\cmss Z\kern-.4em Z}}{\hbox{\cmss Z\kern-.4em Z}}
 {\lower.9pt\hbox{\cmsss Z\kern-.4em Z}}
 {\lower1.2pt\hbox{\cmsss Z\kern-.4em Z}}\else{\cmss Z\kern-.4em Z}\fi}

\begin{document}
\title{
Randomness in the Dark Sector:\\
Emergent Mass Spectra and Dynamical Dark Matter Ensembles
}

\author{Keith R. Dienes$^{1,2}$\footnote{E-mail address:  {\tt dienes@email.arizona.edu}},
      Jacob Fennick$^{3}$\footnote{E-mail address:  {\tt jfennick@hawaii.edu}},
      Jason Kumar$^{3}$\footnote{E-mail address:  {\tt jkumar@hawaii.edu}},
      Brooks Thomas$^{4}$\footnote{E-mail address:  {\tt bthomas@ColoradoCollege.edu}}}
\affiliation{
     $^1\,$Department of Physics, University of Arizona, Tucson, AZ  85721  USA\\
     $^2\,$Department of Physics, University of Maryland, College Park, MD  20742  USA\\
     $^3\,$Department of Physics \& Astronomy, University of Hawaii, Honolulu, HI 96822  USA\\
     $^4\,$Department of Physics, Colorado College, Colorado Springs, CO  80903  USA}

\begin{abstract}
In general, non-minimal models of the dark sector
such as Dynamical Dark Matter posit the existence
of an ensemble of individual dark components
with differing masses, cosmological abundances, and couplings
to the Standard Model.
Perhaps the most critical among these features is the spectrum of masses,
as this goes a long way towards determining the cosmological
abundances and lifetimes of the corresponding states.
Many different underlying theoretical structures can be imagined for the dark sector,
each giving rise to its own mass spectrum and corresponding density of states.
In this paper, by contrast, we investigate the spectrum of masses
that emerges statistically from underlying processes which are essentially
random.
We find a density of states $n(m)$ which decreases as a function of mass and actually
has an upper limit $m_{\rm max}$ beyond which $n(m)=0$.   
We also demonstrate that this ``emergent'' density of states
is particularly auspicious from the perspective of the Dynamical Dark Matter framework,
leading to cosmological abundances and decay widths that are suitably balanced against
each other across the dark-matter ensemble.
Thus randomness in the dark sector coexists quite naturally with Dynamical Dark Matter,
and we examine the prospects for observing
the signals of such scenarios in dark-matter indirect-detection experiments.
\end{abstract}

\maketitle


\newcommand{\PRE}[1]{{#1}} 
\newcommand{\ul}{\underline}
\newcommand{\del}{\partial}
\newcommand{\nbox}{{\,\lower0.9pt\vbox{\hrule \hbox{\vrule height 0.2 cm
\hskip 0.2 cm \vrule height 0.2 cm}\hrule}\,}}

\newcommand{\postscript}[2]{\setlength{\epsfxsize}{#2\hsize}
   \centerline{\epsfbox{#1}}}
\newcommand{\gweak}{g_{\text{weak}}}
\newcommand{\mweak}{m_{\text{weak}}}
\newcommand{\mplanck}{M_{\text{Pl}}}
\newcommand{\mstar}{M_{*}}
\newcommand{\sigmaan}{\sigma_{\text{an}}}
\newcommand{\sigmatot}{\sigma_{\text{tot}}}
\newcommand{\sigmaSI}{\sigma_{\rm SI}}
\newcommand{\sigmaSD}{\sigma_{\rm SD}}
\newcommand{\OmegaM}{\Omega_{\text{M}}}
\newcommand{\OmegaDM}{\Omega_{\text{DM}}}
\newcommand{\ipb}{\text{pb}^{-1}}
\newcommand{\ifb}{\text{fb}^{-1}}
\newcommand{\iab}{\text{ab}^{-1}}
\newcommand{\ev}{\text{eV}}
\newcommand{\kev}{\text{keV}}
\newcommand{\mev}{\text{MeV}}
\newcommand{\gev}{\text{GeV}}
\newcommand{\tev}{\text{TeV}}
\newcommand{\pb}{\text{pb}}
\newcommand{\mb}{\text{mb}}
\newcommand{\cm}{\text{cm}}
\newcommand{\m}{\text{m}}
\newcommand{\km}{\text{km}}
\newcommand{\kg}{\text{kg}}
\newcommand{\g}{\text{g}}
\newcommand{\s}{\text{s}}
\newcommand{\yr}{\text{yr}}
\newcommand{\Mpc}{\text{Mpc}}
\newcommand{\etal}{{\em et al.}}
\newcommand{\eg}{{\em e.g.}}
\newcommand{\ie}{{\em i.e.}}
\newcommand{\ibid}{{\em ibid.}}
\newcommand{\Eqref}[1]{Equation~(\ref{#1})}
\newcommand{\secref}[1]{Sec.~\ref{sec:#1}}
\newcommand{\secsref}[2]{Secs.~\ref{sec:#1} and \ref{sec:#2}}
\newcommand{\Secref}[1]{Section~\ref{sec:#1}}
\newcommand{\appref}[1]{App.~\ref{sec:#1}}
\newcommand{\figref}[1]{Fig.~\ref{fig:#1}}
\newcommand{\figsref}[2]{Figs.~\ref{fig:#1} and \ref{fig:#2}}
\newcommand{\Figref}[1]{Figure~\ref{fig:#1}}
\newcommand{\tableref}[1]{Table~\ref{table:#1}}
\newcommand{\tablesref}[2]{Tables~\ref{table:#1} and \ref{table:#2}}
\newcommand{\met}{\not{\! \! E_T}}
\newcommand{\Dsle}[1]{\slash\hskip -0.28 cm #1}
\newcommand{\Dslp}[1]{\slash\hskip -0.23 cm #1}
\newcommand{\mpt}{{\Dslp p_T}}
\newcommand{\Dsl}[1]{\slash\hskip -0.20 cm #1}

\newcommand{\mB}{m_{B^1}}
\newcommand{\mq}{m_{q^1}}
\newcommand{\mf}{m_{f^1}}
\newcommand{\mKK}{m_{KK}}
\newcommand{\WIMP}{\text{WIMP}}
\newcommand{\SWIMP}{\text{SWIMP}}
\newcommand{\NLSP}{\text{NLSP}}
\newcommand{\LSP}{\text{LSP}}
\newcommand{\mWIMP}{m_{\WIMP}}
\newcommand{\mSWIMP}{m_{\SWIMP}}
\newcommand{\mNLSP}{m_{\NLSP}}
\newcommand{\mchi}{m_{\chi}}
\newcommand{\mgravitino}{m_{\gravitino}}
\newcommand{\mmed}{M_{\text{med}}}
\newcommand{\gravitino}{\tilde{G}}
\newcommand{\Bino}{\tilde{B}}
\newcommand{\photino}{\tilde{\gamma}}
\newcommand{\stau}{\tilde{\tau}}
\newcommand{\slepton}{\tilde{l}}
\newcommand{\snu}{\tilde{\nu}}
\newcommand{\squark}{\tilde{q}}
\newcommand{\mgaugino}{M_{1/2}}
\newcommand{\epsEM}{\varepsilon_{\text{EM}}}
\newcommand{\mmess}{M_{\text{mess}}}
\newcommand{\lmess}{\Lambda}
\newcommand{\nmess}{N_{\text{m}}}
\newcommand{\signmu}{\text{sign}(\mu)}
\newcommand{\Omegachi}{\Omega_{\chi}}
\newcommand{\lambdafs}{\lambda_{\text{FS}}}
\newcommand{\be}{\begin{equation}}
\newcommand{\ee}{\end{equation}}
\newcommand{\bea}{\begin{eqnarray}}
\newcommand{\eea}{\end{eqnarray}}
\newcommand{\baln}{\begin{align}}
\newcommand{\ealn}{\end{align}}
\newcommand{\lsim}{\lower.7ex\hbox{$\;\stackrel{\textstyle<}{\sim}\;$}}
\newcommand{\gsim}{\lower.7ex\hbox{$\;\stackrel{\textstyle>}{\sim}\;$}}

\newcommand{\ssection}[1]{{\em #1.\ }}
\newcommand{\rem}[1]{\textbf{#1}}

\def\ie{{\it i.e.}\/}
\def\eg{{\it e.g.}\/}
\def\etc{{\it etc}.\/}


\section{Introduction\label{sec:Introduction}}


Dynamical Dark Matter~\cite{DDM1,DDM2} (DDM) is an alternative framework
for dark-matter physics in which the dark sector consists of a large
(potentially vast) ensemble of individual dark-matter particles, each
with its own mass, cosmological abundance, and decay width for decays
into Standard-Model (SM) states. 
Within the DDM framework,
the phenomenological viability of such an ensemble is then achieved through
a balancing of lifetimes against cosmological abundances across the entire
set of states comprising the ensemble.
Specifically, states within the DDM ensemble which have greater decay widths
into SM  states must have smaller cosmological abundances,
while states with smaller decay widths may have larger cosmological abundances~\cite{DDM1,DDM2}.
As such, DDM furnishes what may be considered to be the most general
example of a non-minimal dark sector, and even reduces to (and thereby incorporates) the simple
case of a single hyperstable dark-matter particle  
as the number of individual states within the DDM ensemble 
is taken to one.
However, as the number of dark-sector states becomes larger, these ensembles
give rise to rich collider-based, astrophysical, and cosmological
phenomenologies~\cite{DDMAxion,DDMLHC,DDMDD,DDMAMS,DDMComplementarity,CutsAndCorrelations,DDMJeff}
which generalize and even transcend what is possible with a single dark-matter particle alone.

At first glance, it might seem that the masses, abundances and lifetimes
of the individual components of the dark ensemble might be completely arbitrary.
Such a scenario would then require the introduction of a plethora of undetermined
parameters, and would hardly be compelling from a theoretical perspective.
However, as has been discussed in 
prior and ongoing work (see, \eg, Refs.~\cite{DDM1,DDM2,DDMString,DesigningDDMPaper})
the structures of such DDM ensembles 
are {\it not}\/ arbitrary --- they are 
determined according to internal organizing principles which 
describe the entire dark-sector ensemble as a single collective entity,
with the properties of all constituents of the ensemble specified in relation to each other.
As a result, realistic DDM ensembles are characterized by only a
handful of free parameters, and are thus every bit as predictive and as natural
as more traditional dark-matter candidates.

The presence of an organizing principle underpinning the structure of 
the ensemble has been a universal feature of all realistic DDM models to date.  Indeed,
in all cases, such organizing principles 
are then ultimately manifested 
in {\it scaling relations}\/ which 
characterize how the different properties of the ensemble 
constituents scale with respect to each other across the ensemble as a whole.  
There are ultimately three different scaling relations which are critical for the
phenomenology (and eventual viability) of the DDM ensemble.
The first of these is the relationship between the mass of an ensemble constituent 
and its cosmological abundance.  This usually depends on the theoretical structure of the ensemble along 
with the additional choice of a particular 
the additional choice of a particular 
cosmological history.   The second is a relationship between the mass of 
the constituent and its decay width into SM  states.  
This usually depends on the structure of the ensemble along with the additional 
choice of particular couplings between the ensemble and SM  states.
However, there is a third scaling relation which follows directly from the structure
of the ensemble itself, without any additional assumptions.
This is the  relationship between the 
mass of the constituent and the number of constituents with similar masses --- \ie, 
the density of states for the ensemble.  In some sense,
this latter scaling relation can be viewed as the most fundamental,
describing the intrinsic spectral structure of the ensemble 
as a single entity.

There are many possible underlying theoretical structures 
--- \ie, many possible over-arching theoretical constructions ---
that have been shown to give rise to viable 
DDM ensembles.
For example, viable
DDM scenarios are known to exist in which the constituents are
the Kaluza-Klein modes of a single higher-dimensional field~\cite{DDM1,DDM2,DDMAxion}.
Alternatively, viable DDM scenarios exist in which 
the constituents are the ``hadronic'' resonances which appear in the confining phase of a
strongly-coupled theory~\cite{DDMString}, or as the oscillator states of 
a fundamental string~\cite{DDMString}.  In such cases, the sets of
scaling relations which characterize the ensemble are fully determined
by the underlying parameters of the theory.  

In this paper, however, we shall consider another possibility entirely, one in which
the properties of the ensemble constituents are {\it not}\/ dictated by an underlying theoretical construction, but
are instead determined by random dynamical processes in the early universe.  
As we shall demonstrate, this need not be problematic because the required 
scaling relations can actually arise {\it statistically}\/, as {\it emergent} phenomena.
Indeed,
although the mass, decay width, and cosmological abundance of each individual
ensemble constituent continues to be essentially random, 
there exists a well-defined probability 
distribution function (PDF) for each of these quantities across the ensemble 
as a whole.  As the number of constituents in the ensemble grows
large --- the regime of interest for DDM --- the
distribution of each such quantity across the ensemble 
conforms to the corresponding PDF with
an increasingly overwhelming probability.
Thus, such scenarios naturally
give rise to robust --- albeit probabilistic --- predictions for one or
more of the fundamental scaling relations that characterize the ensemble.
Moreover, in scenarios of this sort,
the properties of the ensemble 
continue to be determined --- up to statistical fluctuations --- by only a few model 
parameters.

In this paper we shall provide a concrete example of a scenario in which a 
``statistical'' DDM ensemble of this sort arises.  
This will happen in essentially two steps.
First, we shall demonstrate how random processes can give rise to a
unique mass spectrum for a non-trivial dark sector.
As required, this spectrum can be recast as a scaling relationship between 
the masses and density of states across a dark-sector ensemble.
To do this, we consider a
class of models in which there exists a hidden sector comprising one or more fields which
transform in some non-trivial multiplet representation of a large hidden-sector symmetry
group.  Hidden-sector symmetries of this sort are well-motivated both in grand
unified theories and in string theory.   
If this symmetry remains unbroken, the masses of the individual
component fields within these multiplets --- the fields which play the role of 
the DDM ensemble constituents in models of this sort --- remain equal.  However, 
in cases in which this symmetry group is spontaneously broken, the degeneracy is 
lifted and a non-trivial mass spectrum for these component fields is generated.
This mass spectrum depends on the particular location on the vacuum manifold 
chosen by the symmetry-breaking dynamics.  Since this choice is essentially arbitrary, 
the masses of the individual ensemble constituents are essentially random.  
Nevertheless, as we shall demonstrate, a predictable density-of-states function emerges in the limit that 
the number of component fields is large.  
Moreover, 
we shall find that
this emergent density-of-states function 
is quite unlike those that emerge for KK towers, dark ``hadrons'',
or other theoretical constructs.   Indeed, we shall see that it exhibits classic
hallmarks that reflect its essentially random, statistical origin.

The second step is then to use this statistical scaling relation between masses and density 
of states in order to derive the additional required scaling relations 
involving decay widths/lifetimes and cosmological abundances.
However, this second step need not necessarily involve 
additional random dynamics.
Indeed, once our statistical density-of-states function is specified,
we shall see that standard, well-established mechanisms for deriving 
these additional scaling
relations will suffice. 
In our case, however, we shall nevertheless introduce 
a second (relatively minor) source of randomness into this step as well.
Our purpose in doing this will be to remain as general as possible;
we shall nevertheless find that 
this additional randomness does not disturb the main 
phenomenological features of our construction.

The ultimate result of our construction will be a collection of 
particles whose scaling relations 
are emergent and satisfy the basic criteria for a viable DDM ensemble,
with lifetimes suitably balanced against abundances.
Moreover, as we shall see, our underlying ``statistical'' density-of-states function 
will be responsible in large measure for this success. 
Thus, in this sense, we shall conclude that
randomness in the dark sector coexists quite naturally with DDM.~

This paper is organized as follows.  In Sect.~\ref{sec:Model}, we present a simple
toy model of the sort described above --- \ie, a model
in which our non-minimal dark-sector constituents 
transform as the elements of a multiplet
of a large hidden-sector $SU(N)$ symmetry group,
and in which
a non-trivial mass spectrum for these fields 
is generated via the random spontaneous breaking of this symmetry.
Then, in Sect.~\ref{sec:RandomMatrixGeneration}, 
using the methods of random-matrix theory,  
we analyze the statistical properties of the resulting mass spectrum and the corresponding
density of states.    We stress that our results up to this point are completely general,
and need not have any particular connection to DDM.~  
However, in 
Sect.~\ref{sec:DetectionAbundance}, we then examine how these ensemble constituents 
can be coupled in a self-consistent manner to other fields of the theory, including 
the fields of the SM.~   Such couplings to the SM will then enable us to translate our statistical mass spectra
into statistical relations governing the spectra of decay widths and cosmological abundances,
and we shall find that  
ensembles 
exhibiting the properties of viable DDM ensembles
naturally emerge.
Finally, in Sect.~\ref{sec:IndirectDet}, we investigate the prospects for observing 
the signals of such scenarios in dark-matter indirect-detection experiments.
Our conclusions are summarized in Sect.~\ref{sec:Conclusion}.


\section{Randomness in the Dark Sector:   An Example Model\label{sec:Model}}


In this section, we introduce a simple toy model of the dark sector in which
the mass of each component within the resulting ensemble of states is determined
through an essentially  random process.
Our discussion in this section will be limited to setting up the model itself and the method
by which these masses are generated. 
Sect.~\ref{sec:RandomMatrixGeneration} will then be devoted to an analysis of 
the statistical properties of the resulting mass spectrum and the scaling   
behaviors it exhibits.

Our toy model of the dark sector is as follows.
We begin by considering a scalar field $\phi$ which
transforms in the fundamental representation of some $SU(N)$ symmetry --- a symmetry
which may in principle be either global or local.  The $N$ individual complex
components $\phi_i$ of this $SU(N)$ multiplet will collectively constitute our
ensemble.  In addition to $\phi$, our toy model also includes a real scalar field
$\eta$ which transforms in the adjoint representation of the same $SU(N)$ symmetry.
For notational convenience, we can expand $\eta$ in an $SU(N)$ basis
\begin{equation}
  {\eta} ~\equiv~ \sum_{a=1}^{N^2-1} \eta_a T_a~,
  \label{eq:etaDef}
\end{equation}
where $T_a$ are the generators of $SU(N)$ and
where the fields $\eta_a$ are the corresponding real coefficients in the expansion.
In addition, we assume that
$\phi$ and $\eta$ are charged under distinct $\IZ_2$ symmetries which we call
$\IZ_2^\phi$ and $\IZ_2^\eta$, respectively.  The most general renormalizable scalar
potential for $\phi$ and $\eta$ consistent with these symmetries is
then given by
\begin{eqnarray}
  V_0 & = & \frac{1}{2} M^2 \phi^\dagger \phi
    + \mu^2\mathrm{Tr}[{\eta}{\eta}]
    + \frac{\xi_\phi}{4}(\phi^\dagger\phi)^2
     \nonumber \\ 
    & &
    ~+  \xi_\eta\big(\mathrm{Tr}[{\eta}{\eta}]\big)^2
    + \frac{\xi_1}{2}\phi^\dagger{\eta}{\eta}\phi
    + \xi_2\mathrm{Tr}[{\eta}{\eta}{\eta}{\eta}]~,~~~~~~~
    \label{eq:V0}
\end{eqnarray}
where $M$ and $\mu$ are parameters with dimensions of mass, and where
$\xi_\phi$, $\xi_\eta$, $\xi_1$, and $\xi_2$ are
dimensionless coupling constants.

Let us consider the regime in which $\mu^2 < 0$ and $\xi_\eta > 0$, with
the values of the remaining model parameters such that $\langle \phi \rangle = 0$.
Moreover, for simplicity, let us assume that $\xi_2$ is sufficiently small
that its effects on the vacuum structure of the theory can safely be neglected.
In this regime, the potential in Eq.~(\ref{eq:V0}) is minimized for any set of
vacuum expectation values (VEVs) $v_a \equiv \langle \eta_a \rangle$ which satisfy
the condition
\begin{equation}
  \sum_{a=1}^{N^2-1}v_a^2 ~\equiv~ v^2
    ~=~ -\frac{\mu^2}{2\xi_\eta}~.
  \label{eq:VacuumManifoldCondit}
\end{equation}
Thus, we see that the vacuum manifold for this toy theory is the surface of
an $(N^2-2)$-sphere, and that the $v_a$  can be thought of as the components of an
$(N^2-1)$-dimensional vector with length $v$.  The vacuum-manifold condition
in Eq.~(\ref{eq:VacuumManifoldCondit}) implies that one or more of the $\eta_a$
always acquires a non-zero VEV.  Thus, the $SU(N)$ symmetry is at least partially
broken.  For a generic configuration of VEVs which satisfy this condition, all of
the $v_a$ are non-zero.

Since the vacuum manifold is an equipotential surface,
there is no dynamical principle which determines the direction of this
$(N^2-1)$-dimensional vector in field space.  In the absence of such a
dynamical principle, this direction --- and the corresponding values of the
$v_a$ --- are essentially arbitrary.  {\it It is therefore reasonable to assume
that the particular assignment of $v_a$ values within the vacuum manifold are
determined by random fluctuations in the early universe and that this
assignment itself can therefore also be considered to be effectively
random.}

Although the assignment of $v_a$ values within the vacuum
manifold specified by Eq.~(\ref{eq:VacuumManifoldCondit}) does not affect
$\langle V_0\rangle$, this assignment does have physical consequences.
Chief among the items affected is the spectrum of masses of
the individual components of the $\phi$ multiplet.
The mass matrix $\mathcal{M}^2$ for the individual
components $\phi_i$ of the multiplet $\phi$ receives an additional
contribution from the $v_a$ in the broken phase of the theory.
In particular, the elements of this matrix become
\begin{eqnarray}
  \mathcal{M}^2_{ij} &=& M^2 \mathbb{I}_{ij} +
    \xi_1 (\langle{\eta}\rangle \langle{\eta}\rangle)_{ij} \nonumber \\
    & = & M^2 \mathbb{I}_{ij} +
    \xi_1 \sum_{a=1}^{N^2-1} \sum_{b=1}^{N^2-1} v_a v_b (T_a T_b)_{ij}~,~~
  \label{eq:MassMatrix}
\end{eqnarray}
where $\mathbb{I}$ is the $N\times N$ identity matrix and where
\begin{equation}
  \langle{\eta}\rangle ~\equiv~ \sum_{a=1}^{N^2-1} v_a T_a~
  \label{eq:etaAbsDef}
\end{equation}
is the VEV of the matrix in Eq.~(\ref{eq:etaDef}).  This contribution
lifts the mass degeneracy among the $\phi_i$ and results
in a non-trivial spectrum of masses $m_i$ for the corresponding physical
particles --- a mass spectrum which is sensitive to the particular assignment of
$v_a$ values.  
It is therefore in this way that the randomness of the dark sector affects
the masses of the components of the dark-matter ensemble. 
Indeed, the resulting squared masses $m^2$ of the ensemble constituents
are simply the eigenvalues of the mass matrix in Eq.~(\ref{eq:MassMatrix}), \ie,
\begin{equation}
  m^2 ~=~ M^2 + \xi_1 \lambda^2~,
\label{eq:neweq}
\end{equation}
where $\lambda$ are the eigenvalues of the matrix $\langle {\eta} \rangle$.

We conclude this section with several important comments.
First, we note that while the mass spectrum of the $\phi_i$ is in large part
determined by the random values assigned to the $v_a$, certain properties of the
mass-squared matrix for our ensemble constituents are nevertheless ensured by the
symmetry structure of the theory.  For example, the $\IZ_2^\eta$ symmetry ensures that
the eigenvalues $m_i^2$ of $\mathcal{M}^2$ be positive-definite by forbidding operators
such as $\mathcal{O}_{{\not{\IZ}_2^\eta}} \sim \phi^\dagger \eta \phi$ from appearing in
the scalar potential in Eq.~(\ref{eq:V0}).  This, in turn, ensures that the $\IZ_2^\phi$
symmetry remain unbroken at the minimum of the potential.  Thus, in the absence of
any additional source of $\IZ_2^\phi$-symmetry breaking, the lightest of the $\phi_i$
mass eigenstates is absolutely stable.

It is perhaps also worth emphasizing that no splitting between the masses of the
real and imaginary components of the complex scalars $\phi_i$ results from the
spontaneous breaking of the $SU(N)$ symmetry in this model.
Indeed, even in the broken phase of the theory, the Lagrangian for our toy model
remains invariant under independent phase rotations of each of the $\phi_i$.
This is a reflection of the fact that
even in the case in which all of the $v_a$ are non-zero, the $SU(N)$ symmetry
is not broken completely.  Indeed, the breaking of a symmetry by the VEV of
an adjoint field cannot reduce the rank of the group.  Rather, a residual
$U(1)^{N-1}$ symmetry always remains;  this may be either global or local, depending on
the construction.  Moreover, it can be seen from Eq.~(\ref{eq:V0}) that our theory
is also invariant under an ``accidental'' global $U(1)$ symmetry which corresponds
to an overall phase rotation of the multiplet $\phi$.  This overall $U(1)^{N}$
symmetry corresponds to the invariance of the Lagrangian under phase rotations of
the $N$ different fields $\phi_i$.

Finally, we remark that 
we have yet to specify whether the $SU(N)$ symmetry is local or
global.  Indeed, this choice does not have any effect on the mass spectrum of
the dark-sector ensemble --- at least at tree level.  However, this choice does have other
important phenomenological consequences.  In the case in which this 
symmetry is global, for example, a massless Goldstone boson will appear in the spectrum 
of the low-energy theory for each spontaneously-broken symmetry generator.
The presence of such a large number of Goldstone bosons is difficult to reconcile with 
observational data.  These particles can violate bounds on the number of 
light, thermalized degrees of freedom derived from Cosmic Microwave Background (CMB)
data~\cite{Ade:2015xua}.  They may also potentially mediate long-range interactions 
between dark-matter particles which conflict with bounds from observations of large- 
and small-scale structure (see, \eg, Ref.~\cite{Ackerman:mha}).  There are two ways
of evading these constraints.  One is to gauge the $SU(N)$ symmetry.  In this 
case, the Goldstone bosons are absorbed into the longitudinal polarizations of the 
massive gauge bosons, and no new light degrees of freedom remain in the theory.
The other is to introduce an additional, explicit breaking of the $SU(N)$ symmetry 
which leads to the generation of a small but non-zero mass for each would-be Goldstone 
boson.  Such explicit breaking will have a negligible effect on the mass
spectrum of the ensemble constituents, provided that the associated symmetry-breaking 
terms are small compared to $V_0$.


\section{Randomness in the Dark Sector:  The Emergent Mass Spectrum\label{sec:RandomMatrixGeneration}}


We have seen in the previous section that the mass spectrum of the fields $\phi_i$
in our toy model is sensitive to the effectively random assignment of $v_a$ values
along the vacuum manifold via Eq.~(\ref{eq:neweq}).  
Since these particles
are to play the role of the ensemble constituents, it is critical to understand
how this assignment affects the spectrum of $m_i$ and whether a randomly chosen set
of $v_a$ values consistent with Eq.~(\ref{eq:VacuumManifoldCondit}) can potentially
give rise to an emergent, universal behavior for the corresponding spectrum.

In practice, this randomness can be incorporated into our toy model as follows.
We begin by choosing a set of $N^2-1$ numbers $x_a$ at random from a
Gaussian distribution centered around zero with unit variance.  We then obtain a
set of VEVs $v_a$ for the components of our adjoint field $\eta$ by rescaling these
numbers by a universal constant, chosen such that the vacuum-manifold constraint
in Eq.~(\ref{eq:VacuumManifoldCondit}) is satisfied:
\begin{equation}
  v_a ~=~ v \left(\sum_{b=1}^{N^2-1} x_b^2\right)^{-1/2} x_a~.
\label{eqhow}
\end{equation}
From this set of VEVs, we construct the matrix $\langle{\eta}\rangle$
according to Eq.~(\ref{eq:etaAbsDef}).  This matrix is required by $SU(N)$
invariance to be traceless and Hermitian and required by construction
to satisfy Eq.~(\ref{eq:VacuumManifoldCondit}), but is otherwise a generic
complex $N\times N$ matrix.  Our aim, then, is to study the statistical
properties of an ensemble of such matrices, and in particular the distribution
of their eigenvalues $\lambda_i$.

A substantial literature exists concerning the statistical properties of
ensembles of randomly generated matrices.  Indeed, this is the central
subject of random-matrix theory (for reviews, see, \eg, Ref.~\cite{Mehta}).
We emphasize that the structure of the $\langle{\eta}\rangle$ matrix in our
toy model differs from the canonical matrix structures commonly studied in
the random-matrix literature.  Nevertheless, for purposes of comparison,
it is useful for us to begin our analysis of the properties
our ensemble of $\langle{\eta}\rangle$ matrices with a brief review
of the properties of similar matrix ensembles whose properties are
well documented.

We begin by reviewing the properties of the case of a Gaussian Unitary Ensemble
(GUE) --- a statistical ensemble consisting of $N\times N$ Hermitian matrices $X$.
The off-diagonal elements $X_{ij}$ with $i<j$ of this matrix are complex numbers drawn
from a complex Gaussian distribution centered around zero with variance $v^2/(2N^2)$
for some constant $v$.  Likewise, the diagonal elements $X_{ii}$ (no sum implied)
are real numbers drawn from a standard Gaussian distribution with the same mean
and variance.  The joint probability distribution function (PDF) 
for obtaining a particular set of eigenvalues $\lambda_i$ 
from the GUE is
\begin{equation}
  P_{\mathrm{GUE}}(\lambda_1,...,\lambda_N) ~ = ~
    \frac{1}{Z_N^{(\mathrm{GUE})}}\prod_{k=1}^N e^{-\frac{N^2}{v^2}\lambda_k^2}
    \prod_{i < j} |\lambda_j - \lambda_i|^2~,
  \label{eq:JointPDFGUE}
\end{equation}
where $Z_N^{(\mathrm{GUE})}$ is a normalization constant.  Note that this joint
PDF vanishes whenever two or more of the $\lambda_i$ are equal, a phenomenon called 
eigenvalue repulsion.  The corresponding eigenvalue density --- \ie, the PDF for a
{\it particular}\/ eigenvalue $\lambda$ within this matrix ensemble --- is
obtained by integrating the joint PDF in Eq.~(\ref{eq:JointPDFGUE}) over the
remaining $\lambda_i$.  The result is
\begin{eqnarray}
  P_{\mathrm{GUE}}(\lambda) & = &
    \frac{e^{-\frac{N^2}{v^2}\lambda^2}}
      {\sqrt{\pi}\,v}\sum_{k=0}^{N-1}
    \frac{H_k^2\big(\frac{N\lambda}{v}\big)}{2^k k!}~,
  \label{eq:OneEigPDFGUE}
\end{eqnarray}
where $H_k(x)$ denotes the Hermite polynomial of degree $k$.

While our ensemble of $\langle{\eta}\rangle$ is closely related to the GUE,
it is even more closely related to a class of matrix ensembles known as
fixed-trace ensembles (FTE).  In particular,
this class of ensembles includes ensembles of complex, Hermitian matrices $X$
which are generated in the same manner as those of the GUE, but subject to
the additional constraint $\mathrm{Tr}[X^\dagger X] = v^2/2$ for some constant $v$.
The joint PDF for the eigenvalues of such an 
FTE is~\cite{LloydPagels,ZyczkowskiSommers}
\begin{eqnarray}
  P_{\mathrm{FTE}}(\lambda_1,...,\lambda_N) & = &
    \frac{1}{Z_N^{(\mathrm{FTE})}}
    \delta\left(\frac{v^2}{2}- \sum_{\ell=1}^N \lambda_\ell^2\right) ~~~~~~~~~~~
    \nonumber \\ & & \times 
    \prod_{i < j} |\lambda_j - \lambda_i|^2~,
  \label{eq:JointPDFFTE}
\end{eqnarray}
where $\delta(x)$ denotes the Dirac $\delta$-function  and where
$Z_N^{(\mathrm{FTE})}$ is a normalization constant, which is in general
different from $Z_N^{(\mathrm{GUE})}$.  
Note that this PDF vanishes if $|\lambda_i| > v / \sqrt{2}$ for any $\lambda_i$,
as the $\delta$-function constraint can no longer be satisfied
in such cases.
As with the GUE, an
analytic expression for the eigenvalue density within the FTE can be
obtained by integrating this joint PDF over the remaining $\lambda_i$.
In particular, 
for $\lambda$ within the range $|\lambda| < v / \sqrt{2}$,
we find~\cite{Akemann,Delannay}
\begin{eqnarray}
  P_{\mathrm{FTE}}(\lambda) & = &
    \frac{\sqrt{2}\Gamma\big({\textstyle\frac{N^2}{2}}\big)}
      {N v\pi\Gamma\left(\frac{N^2 - 1}{2}\right)}
    \sum_{j=0}^{N-1}\frac{(-2)^j\Gamma\left(j + {\textstyle\frac{1}{2}}\right)}{j!}
    \left(\!\!\begin{array}{c} N \\ j + 1\end{array}\!\!\right)
    \nonumber \\ & &
      \times \,
      \rule[0pt]{0pt}{11pt}_2F_1\!\left(j + {\textstyle\frac{1}{2}},
      {\textstyle\frac{3 - N^2}{2}};
      {\textstyle\frac{1}{2}};
      {\textstyle\frac{2\lambda^2}{v^2}}\right)~,
   \label{eq:OneEigPDFFTE}
\end{eqnarray}
where $\Gamma(x)$ is
the Euler gamma function and where $_2F_1\left(a, b; c; x\right)$ is
the ordinary hypergeometric function.
By contrast, $P_{\mathrm{FTE}}(\lambda) = 0$ 
for all $|\lambda| > v / \sqrt{2}$.

Had we chosen the symmetry group in our toy model to be $U(N)$ rather than
$SU(N)$, our $\langle{\eta}\rangle$ matrices would be the members
of precisely this fixed-trace ensemble.  By contrast, for the $SU(N)$ case,
not only does Eq.~(\ref{eq:VacuumManifoldCondit}) impose a constraint
\begin{equation}
  \mathrm{Tr}\big[\langle{\eta}\rangle\langle{\eta}\rangle\big] ~=~
  \sum_{i=1}^N \lambda_i^2 ~=~ \frac{v^2}{2}
\end{equation}
on the squares of the eigenvalues $\lambda_i$, but the tracelessness condition
on the $SU(N)$ generators $T_a$ imposes an additional constraint
\begin{equation}
  \mathrm{Tr}\big[\langle{\eta}\rangle\big] ~=~ \sum_{i=1}^N \lambda_i ~=~ 0~
  \label{eq:TracelessnessConstraint}
\end{equation}
on the eigenvalues themselves.  Thus, the joint PDF for our ensemble of
$\langle{\eta}\rangle$ matrices takes the form
\begin{eqnarray}
  P_{\rm SU(N)}(\lambda_1,...,\lambda_N) & = &
    \frac{1}{Z_N^{(\eta)}}
    \delta\bigg(\sum_{p=1}^N \lambda_p\bigg) 
    \delta\bigg(\frac{v^2}{2}- \sum_{\ell=1}^N \lambda_\ell^2\bigg) \nonumber\\
    & & \times\, 
    \prod_{i < j} |\lambda_j - \lambda_i|^2~.
  \label{eq:JointPDFEta}
\end{eqnarray}

To the best of our knowledge, no analytic expression for the corresponding 
eigenvalue density $P_{\rm SU(N)}(\lambda)$ for arbitrary $N$ currently exists 
in the literature.  However, as we shall see, the eigenvalue densities which 
we obtain numerically for matrices drawn from this ensemble bear many similarities 
to $P_{\mathrm{GUE}}(\lambda)$ and $P_{\mathrm{FTE}}(\lambda)$ for the same value 
of $N$.  Perhaps the most important of these similarities
is that $P_{\mathrm{GUE}}(\lambda)$, $P_{\mathrm{FTE}}(\lambda)$, and
$P_{\rm SU(N)}(\lambda)$ all begin to coincide for $N\gg 1$.
In particular, for $N\gg 1$,
all three of these eigenvalue-density functions converge
to the Wigner semicircle distribution~\cite{Wigner}
\begin{equation}
  P_{\mathrm{WS}}(\lambda) ~ = ~
 \begin{cases}
   \displaystyle{\frac{\sqrt{2N}}{\pi v}\sqrt{1-\frac{N\lambda^2}{2v^2}}}
   &     {\rm for} ~|\lambda| < v \sqrt{2/N} \\
    0 & {\rm otherwise}~.
\end{cases}
\label{eq:OneEigPDFWigner}
\end{equation}

Given the relationship in Eq.~(\ref{eq:neweq}) between the eigenvalues $\lambda$
and the corresponding ensemble masses $m$,
it is relatively straightforward to 
convert the eigenvalue-density function $P(\lambda)$ associated with any
matrix ensemble into a corresponding density-of-states function $n(m)$ 
for the mass-eigenstate fields $\phi_i$.
In general, the density-of-states function $n(m)$ with respect to the ensemble-constituent
mass $m$ is simply the product of the number of states  in the ensemble
 and the PDF for that value of $m$ within the ensemble:
\begin{equation}
           n(m) ~\equiv~ N \,P(m)~.
\end{equation}
The PDF $P(m)$ can be
obtained directly from the corresponding eigenvalue density $P(\lambda)$ via a
change of variables.  It is more convenient, however, for us to express the
density of states of ensemble constituents as a function of the dimensionless
mass variable $\tilde{m} \equiv \sqrt{2} m /(\sqrt{\xi_1}v)$ rather than the mass
$m$ itself.  We also define the similarly rescaled mass parameter
$\tilde{M} \equiv \sqrt{2} M/(\sqrt{\xi_1}v)$.
The density-of-states function $n (\tilde{m})$ with respect to $\tilde{m}$ is
then given by
\begin{equation}
  n (\tilde{m}) ~ = ~ 2 N P (\lambda)\frac{d\lambda}{d\tilde{m}}
    ~=~ \frac{\sqrt{2} v N \tilde{m}}{\sqrt{\tilde{m}^2 - \tilde{M}^2}}
    P (\lambda)~,
  \label{eq:DOSFunctionEta}
\end{equation}
where the factor of two in this expression arises due to the fact that
$\lambda^2$ is positive-definite.  The corresponding density-of-states
function $n (m)$ with respect to the constituent mass $m$ rather than
the dimensionless variable $\tilde{m}$ is then given by
\begin{equation}
  n (m) ~=~ \frac{\sqrt{2}n (\tilde{m})}{\sqrt{\xi_1}v}~.
\end{equation}

Without an analytic expression for $P_{\rm SU(N)}(\lambda)$, we cannot write down
a closed-form expression for $n_{\rm SU(N)}(\tilde{m})$ for arbitrary $N$.
Nevertheless, for $N\gg 1$ (where the Wigner semicircle approximation holds),
we can obtain an analytic approximation for $n_{\rm SU(N)}(\tilde{m})$ by taking 
$P_{\rm SU(N)}(\lambda)\approx P_{\mathrm{WS}}(\lambda)$ in Eq.~(\ref{eq:DOSFunctionEta}).
Moreover, we can gain insight into the behavior of $n_{\rm SU(N)}(\tilde{m})$ for smaller
values of $N$ by examining the properties of the density-of-states functions 
$n_{\mathrm{GUE}}(\tilde{m})$ and $n_{\mathrm{FTE}}(\tilde{m})$ which one would obtain 
for a GUE and FTE, respectively.  Indeed, exact analytic expressions for these functions 
can be obtained for arbitrary $N$ by substituting the eigenvalue densities in
Eqs.~(\ref{eq:OneEigPDFGUE}) and~(\ref{eq:OneEigPDFFTE}) into 
Eq.~(\ref{eq:DOSFunctionEta}).  These analytic expressions, as well as the 
Wigner semicircle expression to which $n_{\mathrm{GUE}}(\tilde{m})$, 
$n_{\mathrm{FTE}}(\tilde{m})$, and $n_{\rm SU(N)}(\tilde{m})$ all tend for $N\gg 1$,
are given below:
\begin{widetext}
\begin{eqnarray}
  n_{\mathrm{GUE}}(\tilde{m}) &=&
    \begin{cases}
      \displaystyle \frac{\sqrt{2}\tilde{m}Ne^{-\frac{1}{2} N^2(\tilde{m}^2-\tilde{M}^2)}}
         {\sqrt{\pi(\tilde{m}^2-\tilde{M}^2)}} 
         \sum_{k=0}^{N-1}
         \frac{H_k^2\left(N\sqrt{\frac{\tilde{m}^2-\tilde{M}^2}{2}}\right)}{2^k k!} ~~~~&
         \tilde{M} \leq  \tilde{m} ~~~~
                      \\
      0 & \mbox{otherwise}~
    \end{cases} \nonumber \\
 n_{\mathrm{FTE}}(\tilde{m}) &=&
   \begin{cases}
   \displaystyle
   \frac{2\tilde{m}\Gamma\big({\textstyle\frac{N^2}{2}}\big)}
     {\pi\sqrt{\tilde{m}^2-\tilde{M}^2}}
   \sum_{j=0}^{N-1}\frac{(-2)^j\Gamma\left(j + {\textstyle\frac{1}{2}}\right)}
     {j!\Gamma\left(\frac{N^2 - 1}{2}\right)}
     \left(\!\!\begin{array}{c} N \\ j + 1\end{array}\!\!\right)
     \rule[0pt]{0pt}{11pt}_2F_1\!\left(j + {\textstyle\frac{1}{2}},
     {\textstyle\frac{3 - N^2}{2}};
     {\textstyle\frac{1}{2}};
     {\textstyle \tilde{m}^2 - \tilde{M}^2}\right)~~~~
     & \tilde{M} \leq  \tilde{m} < \displaystyle \sqrt{\tilde{M}^2 + 1 } \\
   0 & \mbox{otherwise}~
    \end{cases} \nonumber \\
 n_{\mathrm{WS}}(\tilde{m}) &=&
  \begin{cases}
   \displaystyle \frac{2 N^{3/2}}{\pi}
     \sqrt{\frac{\tilde{m}^2}{\tilde{m}^2-\tilde{M}^2}-\frac{N\tilde{m}^2}{4}}~~~~
     & \tilde{M} \leq  \tilde{m} < \displaystyle \sqrt{\tilde{M}^2 + \frac{4}{N}} \\
   0 & \mbox{otherwise}~.
  \end{cases}
  \label{eq:DensityOfStateswithm}
\end{eqnarray}
\end{widetext}

\begin{figure}[h]
  \includegraphics[width=0.4\textwidth, keepaspectratio]{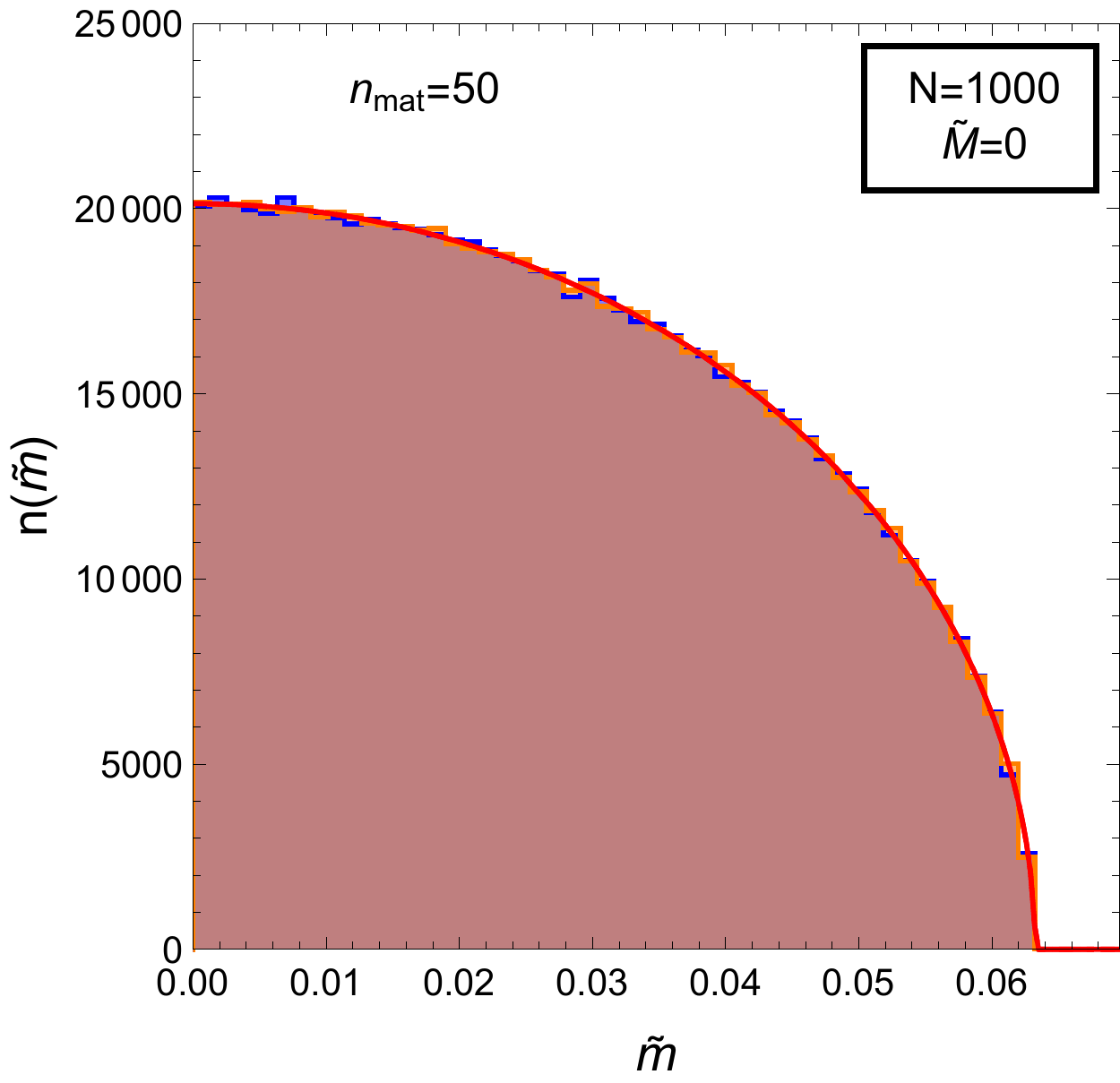}
  \caption{The density-of-states function $n(m)$ for the ensemble
    constituents $\phi_i$ for $N=1000$.  What initially might appear as a single
    brownish histogram actually represents the superposition of two nearly identical
    histograms, one blue and one orange, each of which
    represents a data set comprising the ensemble-constituent masses
    obtained for a sample of $n_{\mathrm{mat}} = 50$ randomly generated
    $\langle{\eta}\rangle$ matrices.  The blue histogram corresponds
    to a model in which the symmetry group is $SU(N)$, while the orange histogram
    corresponds to a model in which the symmetry group is $U(N)$.
    Likewise, the red curve represents the expression for $n_{\mathrm{WS}}(\tilde{m})$ in
    Eq.~(\protect\ref{eq:DensityOfStateswithm}).  As discussed in the text, 
      for such large values of $N$ 
    as assumed here both histograms indeed coincide 
       and are essentially indistinguishable from
    $n_{\mathrm{WS}}(\tilde{m})$. 
  \label{fig:DOSFunctionN1000}}
\end{figure}

\begin{figure*}[t]
  \includegraphics[width=0.3\textwidth, keepaspectratio]{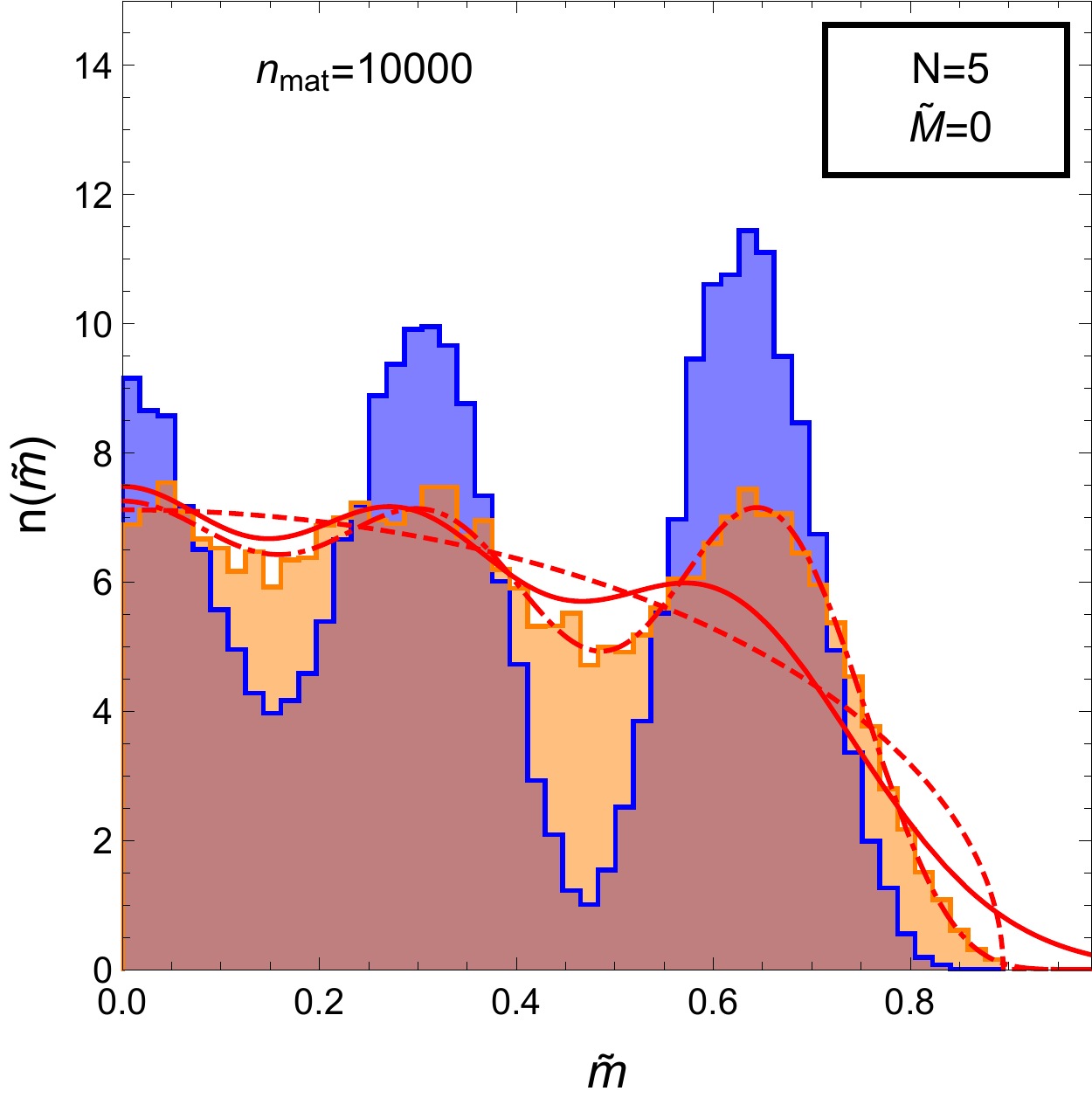}~~~
  \includegraphics[width=0.3\textwidth, keepaspectratio]{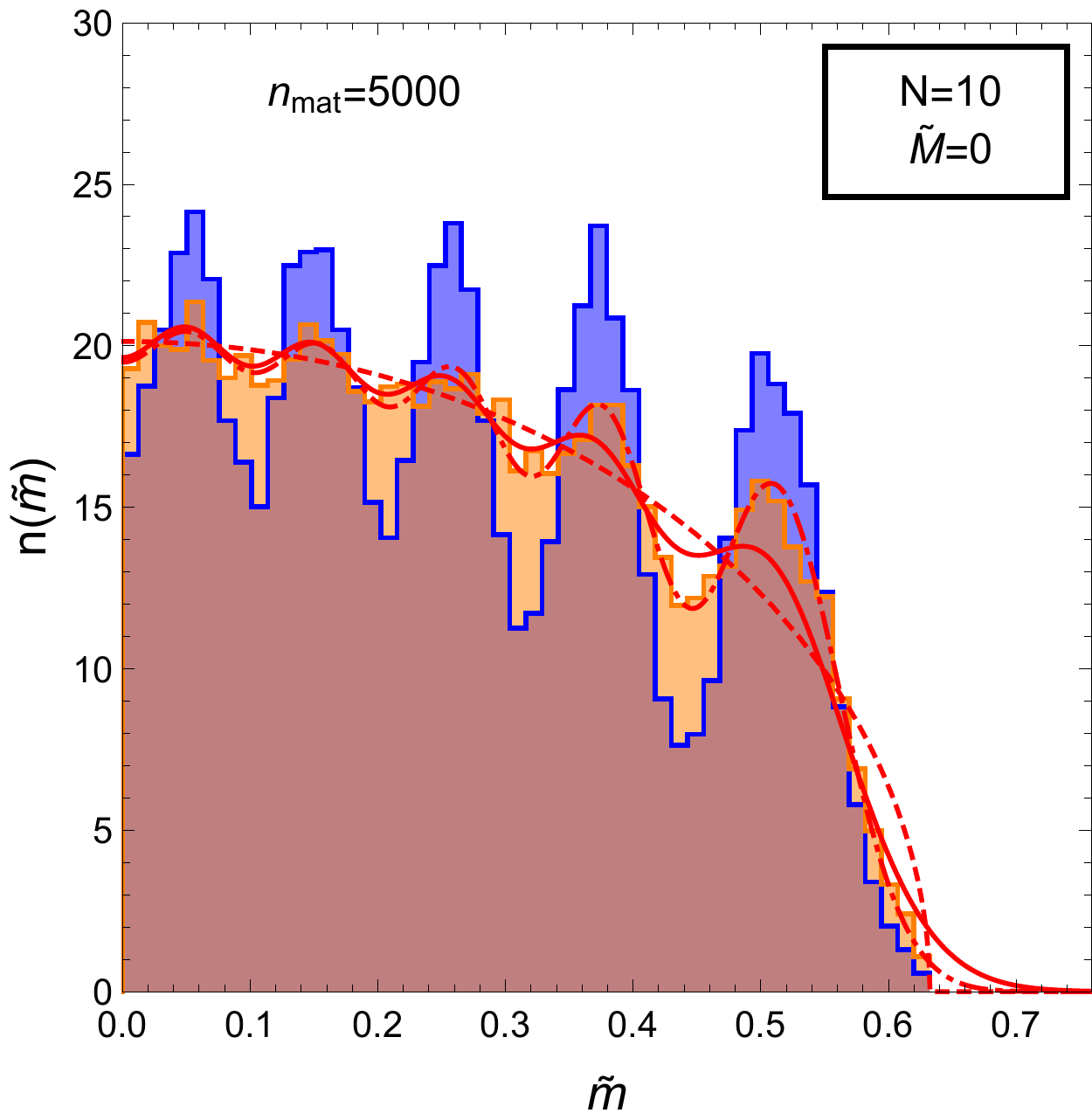}~~~
  \includegraphics[width=0.3\textwidth, keepaspectratio]{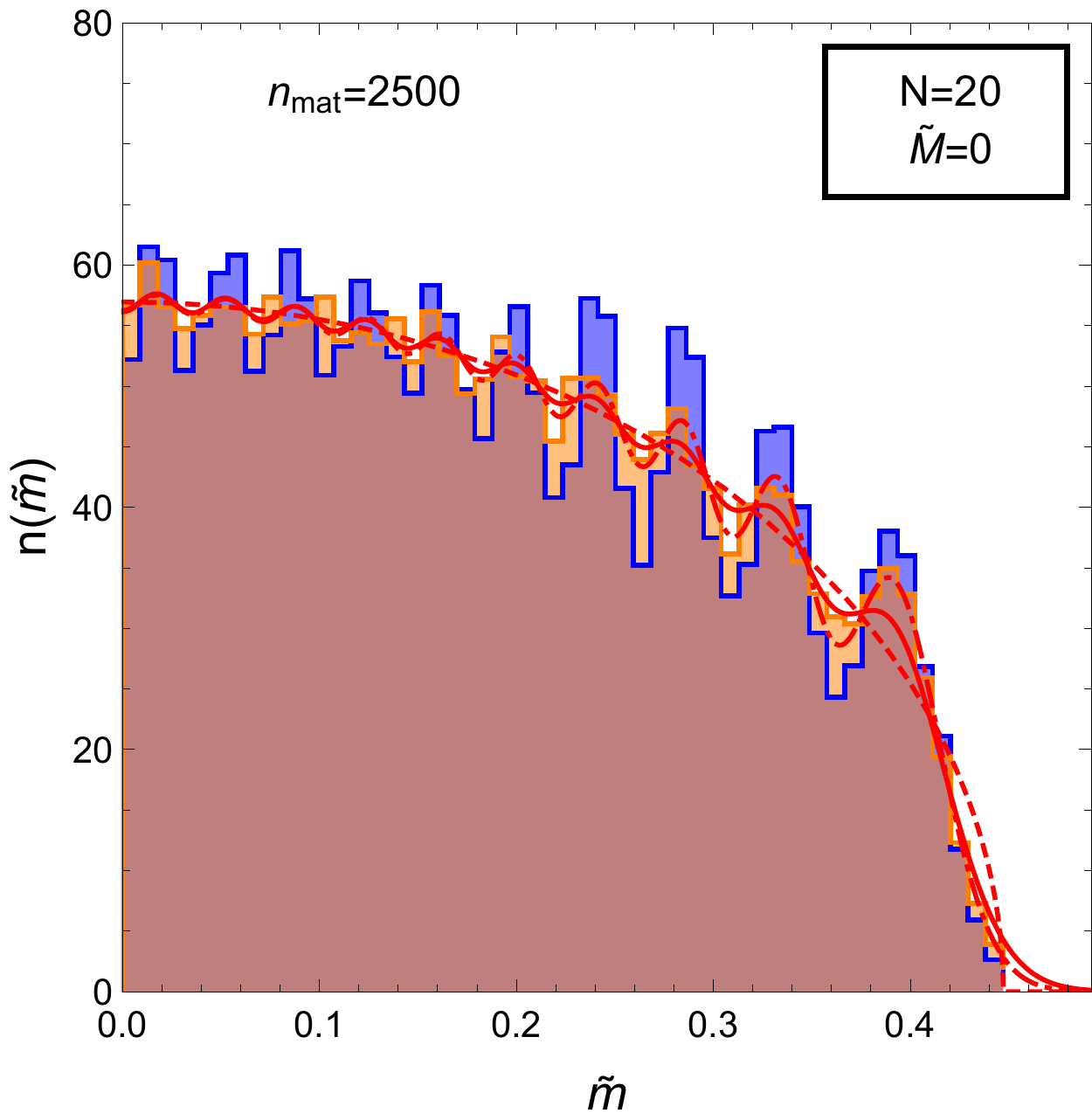}
  \caption{The density-of-states function $n(m)$ for the ensemble
    constituents $\phi_i$ for $N=5$ (left panel), $N=10$ (center panel), and
    $N=20$ (right panel).  In each panel, the blue and orange histograms
    each represent a data set comprising the ensemble-constituent masses
    obtained for a sample of $n_{\mathrm{mat}}$ randomly generated
    $\langle{\eta}\rangle$ matrices, where $n_{\mathrm{mat}}$ is chosen
    in each case such that the total number of individual mass eigenvalues
    in the data set is $5 \times 10^4$.  As in Fig.~\protect\ref{fig:DOSFunctionN1000},
    the blue histogram corresponds to a model in which the symmetry group is
    $SU(N)$, while the orange histogram corresponds to a model in which the symmetry
    group is $U(N)$.  The solid, dot-dashed, and dashed curves in
    each panel correspond to the functions $n_{\mathrm{GUE}}(\tilde{m})$,
    $n_{\mathrm{FTE}}(\tilde{m})$, and $n_{\mathrm{WS}}(\tilde{m})$ in
    Eq.~(\protect\ref{eq:DensityOfStateswithm}), respectively.
  \label{fig:DOSFunctionSmallN}}
\end{figure*}

With these analytic expressions as a guide, we now numerically examine the 
distributions of masses obtained for statistical ensembles based on an $SU(N)$ 
symmetry group.  In particular, we perform a numerical analysis 
of these distributions according to the following procedure.
For a given value of $N$, we randomly generate a number $n_{\mathrm{mat}}$
of $\langle{\eta}\rangle$ matrices
following the procedure described above Eq.~(\ref{eqhow}),
where $n_{\mathrm{mat}}$ is chosen
such that the total number of individual eigenvalues in the data set
is fixed to a reference value which we take to be  $5 \times 10^4$.  
For each of these matrices, we diagonalize the
corresponding mass-squared matrix for the ensemble constituents in order to 
obtain their masses $m_i$.  We then combine the $m_i$ values for
all of the matrices in our sample into a single data set, a histogram of
which provides a numerical approximation to the corresponding analytic 
density-of-states function $n_{\rm SU(N)}(\tilde{m})$.  For purposes of comparison,
we also derive a numerical approximation for $n_{\mathrm{FTE}}(\tilde{m})$ --- 
the density-of-states function obtained for a $U(N)$ rather than an 
$SU(N)$ symmetry group --- using the same procedure.

In Fig.~\ref{fig:DOSFunctionN1000}, we show the density-of-states function
$n(\tilde{m})$ for the ensemble constituents generated by diagonalizing
randomly generated mass matrices, for $N=1000$ and $\tilde{M}=0$.
The blue histogram corresponds to the $SU(N)$ case, while
the orange histogram corresponds to the $U(N)$ case.
In each case, the results correspond to a data set
comprising the ensemble-constituent masses obtained for a sample of
$n_{\mathrm{mat}} = 50$ randomly generated matrices.
By contrast, the red curve in Fig.~\ref{fig:DOSFunctionN1000}
represents the analytical function
$n_{\mathrm{WS}}(\tilde{m})$ in Eq.~(\ref{eq:DensityOfStateswithm}).  
As evident from Fig.~\ref{fig:DOSFunctionN1000}, the function 
$n_{\mathrm{WS}}(\tilde{m})$ provides an excellent approximation 
to the histogram data, as claimed.  Moreover, for such a large value of $N$, 
the functions $n_{\mathrm{GUE}}(\tilde{m})$ and $n_{\mathrm{FTE}}(\tilde{m})$
are essentially indistinguishable from $n_{\mathrm{WS}}(\tilde{m})$.
Thus all three functions do an excellent job of approximating the data for $N\gg 1$.

It is also interesting to examine how the results are modified in the regime
in which $N$ is small and the Wigner semicircle distribution provides a less
reliable approximation for the true density of states for the ensemble.
In Fig.~\ref{fig:DOSFunctionSmallN}, we show the corresponding results for
$N = 5$ (left panel), $N = 10$ (center panel), and $N=20$ (right panel).
In each panel, the blue and orange histograms each represent a data set
comprising the ensemble-constituent masses obtained for a sample of
$n_{\mathrm{mat}}$ randomly generated $\langle{\eta}\rangle$ matrices,
where $n_{\mathrm{mat}}$ is chosen in each case such that the total number
of individual mass eigenvalues in the data set is $5 \times 10^4$.
As in Fig.~\protect\ref{fig:DOSFunctionN1000}, the blue histogram corresponds
to a model in which the symmetry group is $SU(N)$ while the orange histogram corresponds
to a model in which the symmetry group is $U(N)$.
The solid, dot-dashed, and dashed curves in each panel
correspond to $n_{\mathrm{GUE}}(\tilde{m})$, $n_{\mathrm{FTE}}(\tilde{m})$, and
$n_{\mathrm{WS}}(\tilde{m})$, respectively.

We see from Fig.~\ref{fig:DOSFunctionSmallN}
that $n_{\mathrm{WS}}(\tilde{m})$ aptly characterizes the overall
``envelope'' of the true density of states for small $N$ in both the $U(N)$
and $SU(N)$ cases, just as it does for large $N$.  However, for small $N$, the
true density of states for these cases also exhibits
 {\it oscillations}\/ around this envelope --- oscillations which grow increasingly
pronounced with decreasing $N$.  For the case in which the symmetry group is
$U(N)$, the histogram data are distributed according to the density-of-states
function $n_{\mathrm{FTE}}(\tilde{m})$, as expected; 
this function actually includes the oscillations, and the histogram differs 
from  $n_{\mathrm{FTE}}(\tilde{m})$ only because of random fluctuations.
However, for the case in which the symmetry group is $SU(N)$, the amplitude of 
the oscillations is significantly more pronounced, and we see that neither 
$n_{\mathrm{FTE}}(\tilde{m})$ nor $n_{\mathrm{WS}}(\tilde{m})$ are accurate 
descriptions of the density of states for the case of an $SU(N)$ symmetry group at 
small $N$.   However, as $N$ grows larger (as illustrated in the right panel of Fig.~\ref{fig:DOSFunctionSmallN}),
the oscillations fade away in relative magnitude for both the $U(N)$ and $SU(N)$ cases, and 
the actual distributions qualitatively begin to approach 
$n_{\mathrm{WS}}(\tilde{m})$.

\begin{figure}[t]
  \includegraphics[width=0.4\textwidth, keepaspectratio]{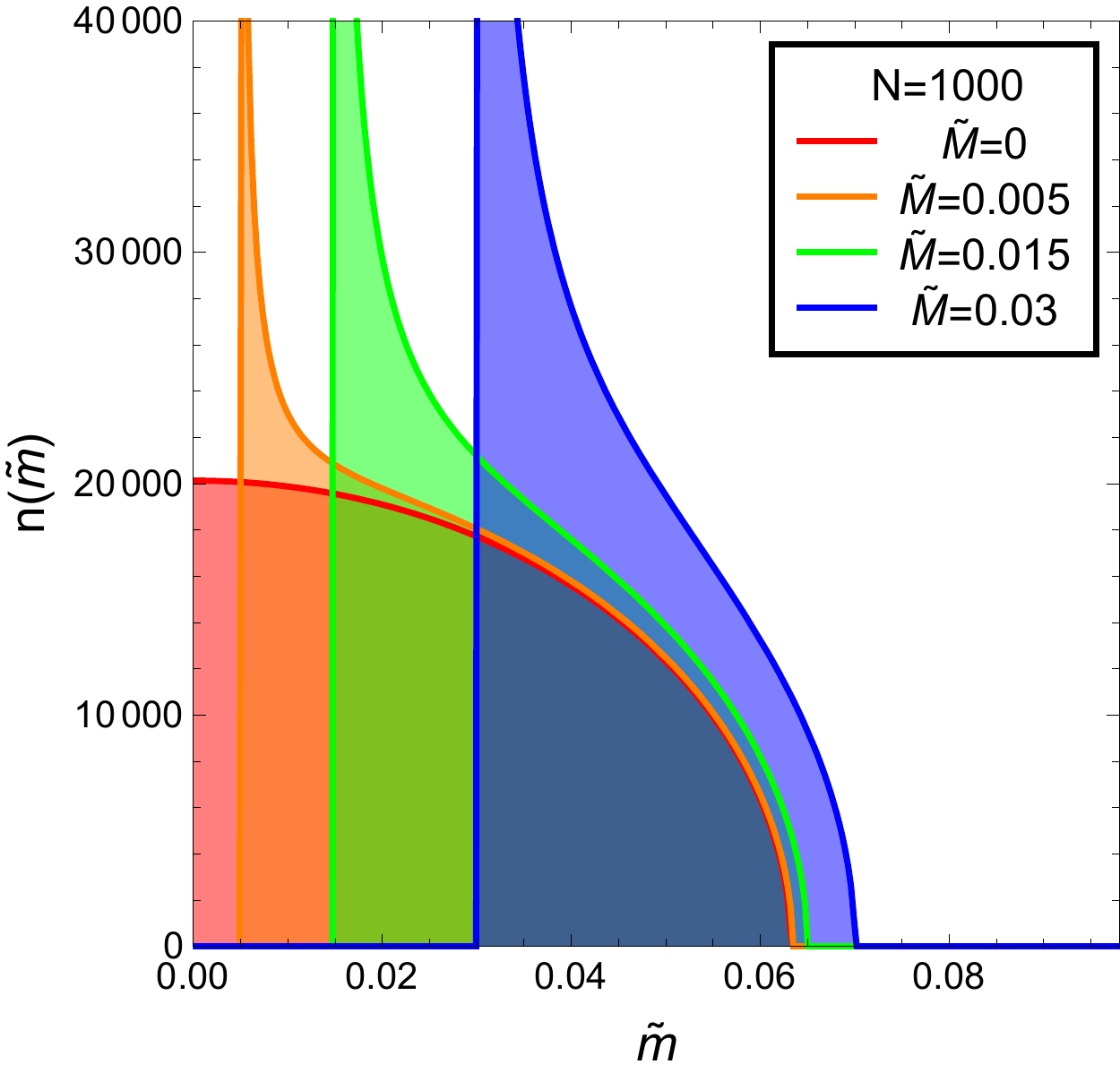}
  \caption{The density-of-states function $n_{\mathrm{WS}}(\tilde{m})$ for
    several different choices of the mass parameter $\tilde{M}$.  In each case,
    we have taken $N = 1000$ and $n_{\rm mat}=50$.
  \label{fig:DOSFunctionMadMLargeN}}
\end{figure}

\begin{figure*}[t]
  \includegraphics[width=0.3\textwidth, keepaspectratio]{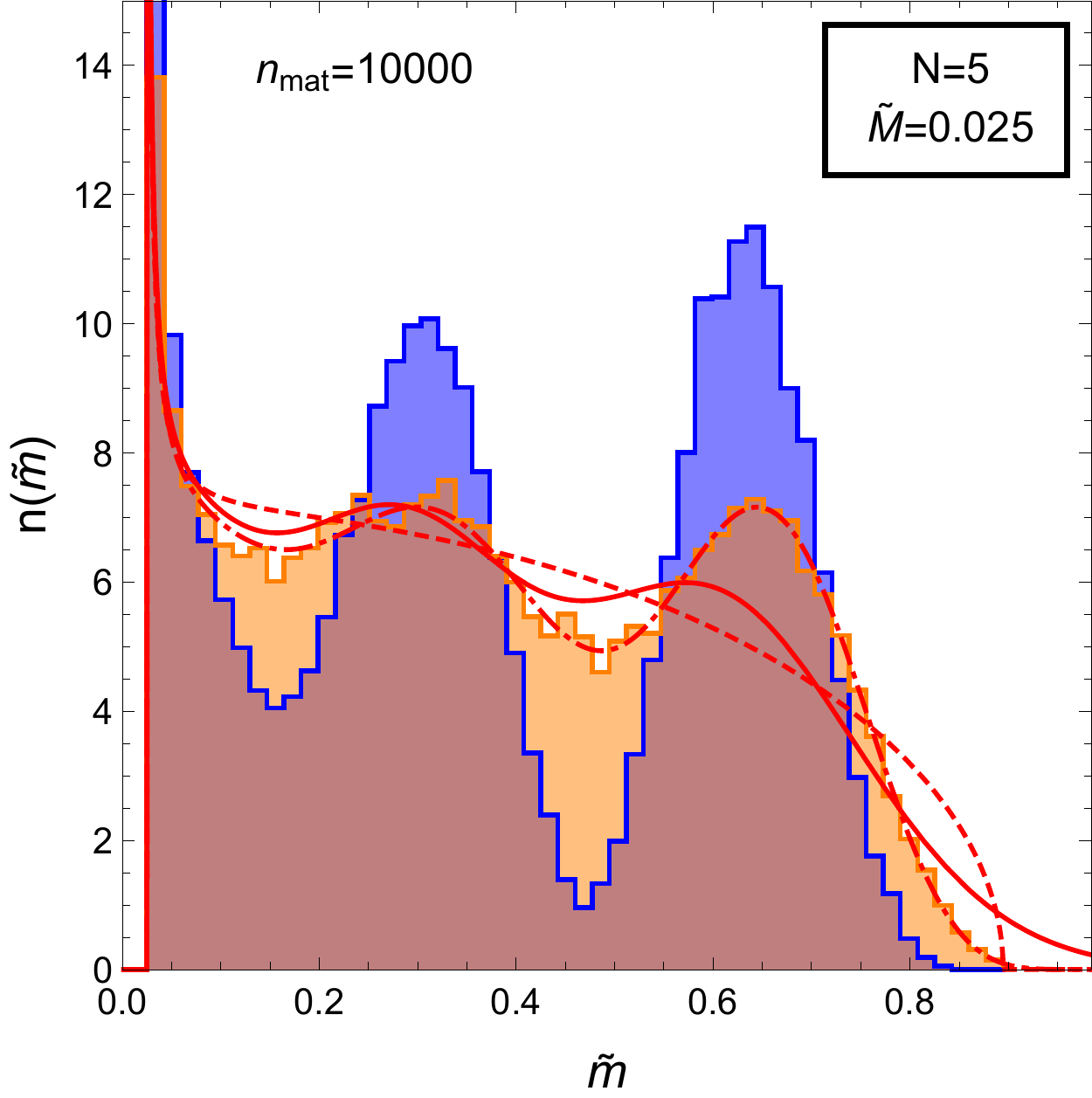}~~~
  \includegraphics[width=0.3\textwidth, keepaspectratio]{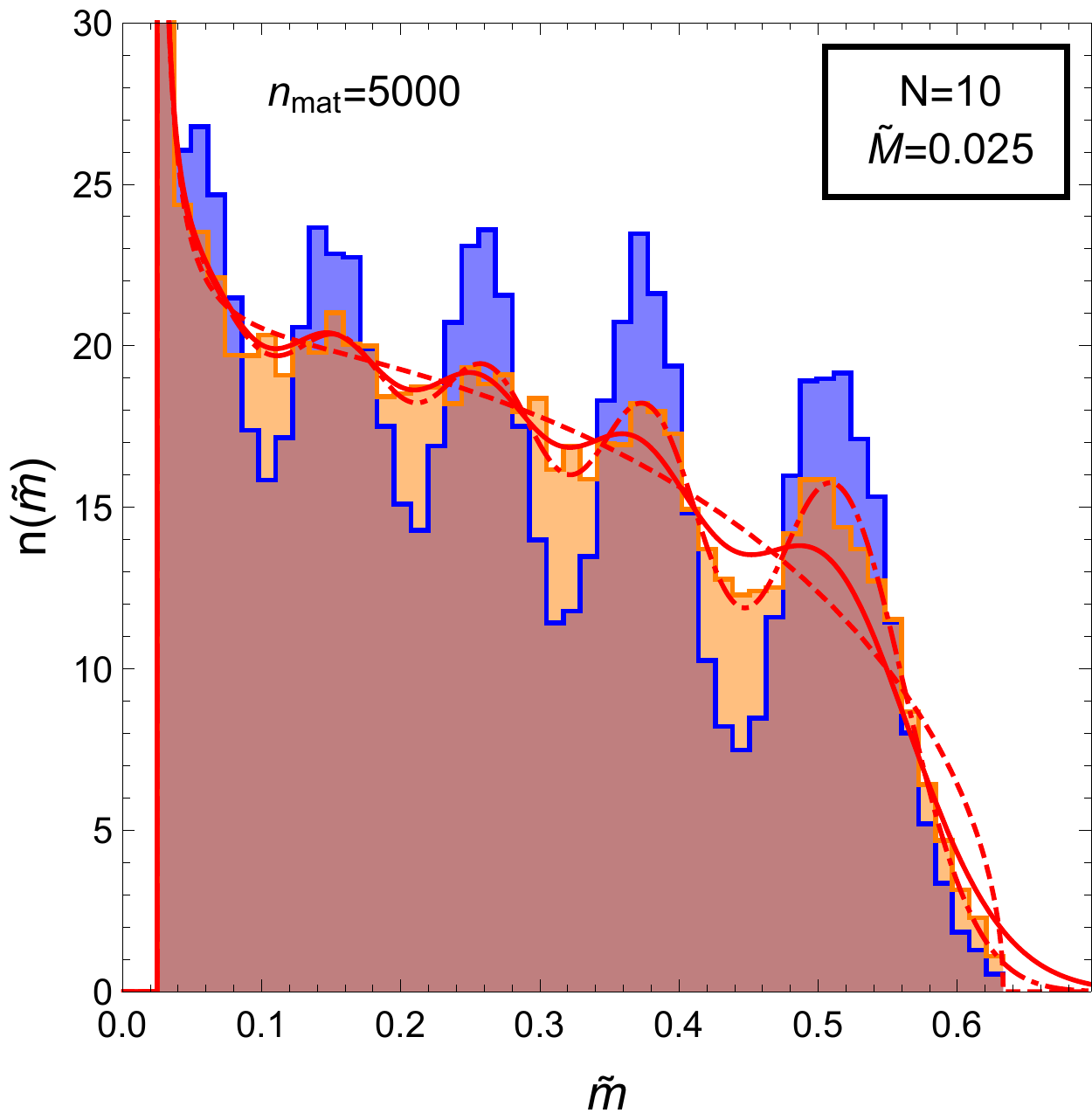}~~~
  \includegraphics[width=0.3\textwidth, keepaspectratio]{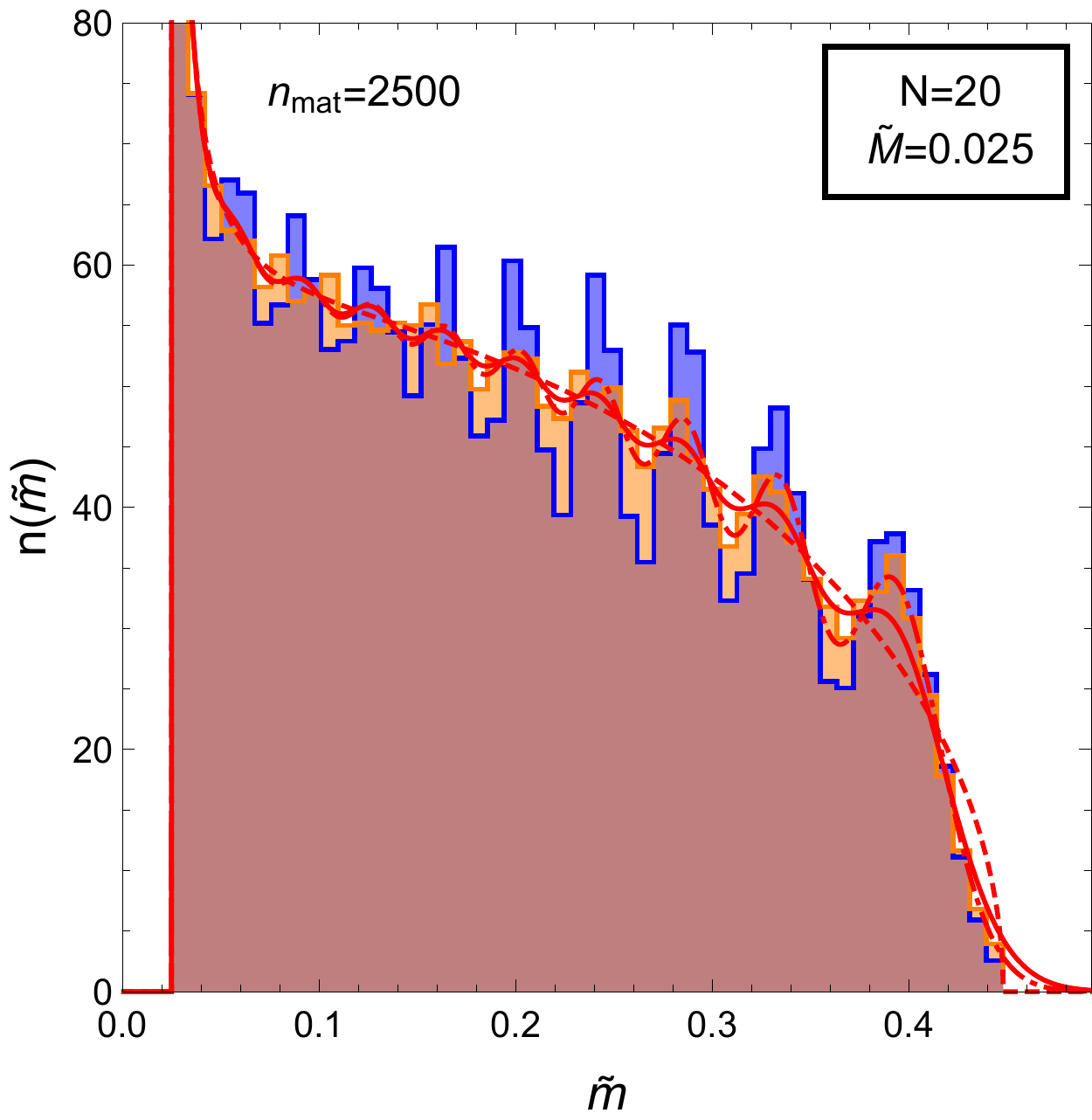}
  \caption{Same as Fig.~\protect\ref{fig:DOSFunctionSmallN},
  but with $\tilde{M}=0.025$ rather than $\tilde{M} = 0$.
  \label{fig:DOSFunctionMadMSmallN}}
\end{figure*}

We now turn to consider how the density of states for our ensemble depends
on $\tilde{M}$.  In Fig.~\ref{fig:DOSFunctionMadMLargeN}, we display curves of
the Wigner density-of-states function $n_{\mathrm{WS}}(\tilde{m})$ for different
values of $\tilde{M}$.  For each of these curves, we have taken
$N=1000$ --- a suitably large value of $N$ for which this function provides an
excellent approximation to the exact density-of-states function $n(\tilde{m})$, in
the case of either a $U(N)$ or $SU(N)$ symmetry group.
We see that for $\tilde{M} \neq 0$, the presence of the additional contribution
to the mass matrix in Eq.~(\ref{eq:MassMatrix}) distorts the density-of-states
function away from the semicircle form which arises in the $\tilde{M} = 0$ case.
This is simply a reflection of the fact that states $\phi_i$ in the ensemble which
receive comparatively small contributions $\xi_1\lambda_i^2 \lesssim M^2$ to their
squared masses from symmetry breaking will have masses $m_i \approx M$.

In Fig.~\ref{fig:DOSFunctionMadMSmallN}, we show the corresponding densities of states
(similar to Fig.~\ref{fig:DOSFunctionSmallN}) for smaller values of $N$ but non-zero 
$\tilde M$.  These results exhibit the same qualitative behavior as in
Fig.~\ref{fig:DOSFunctionSmallN}.

The main results of this section thus far can be summarized as follows.  We have shown that
in scenarios in which the ensemble constituents are the component fields of
a multiplet which transforms under a spontaneously-broken symmetry group, the 
PDF for the masses of those constituents takes a well-defined form.
As the number of component fields in the multiplet increases, the density-of-states 
function $n_{\rm SU(N)}(m)$ for our ensemble becomes overwhelmingly statistically likely 
to coincide with that PDF.~  Moreover, the properties of this PDF function for the density of states 
are also significant.  In particular we observe that the density-of-states function  
generally {\it decreases}\/ with increasing $m$, and does so in a predictable way with
a well-defined upper limit $n_{\rm max}$ beyond which $n(m)=0$.
By contrast, other natural scenarios which naturally give rise to large numbers of
dark particles, such as Kaluza-Klein towers~\cite{DDM1,DDM2}, bound-state resonances 
in strongly coupled theories~\cite{DDMString}, or oscillator states of a fundamental
string~\cite{DDMString} give rise to density-of-state functions which are either
independent of $m$ or else rise exponentially with $m$.  

The emergent mass spectrum we have found in this section
is ultimately the hallmark of the underlying randomness within our toy model.
Indeed, this behavior for the emergent mass spectrum ---
and the Wigner semi-circle rule from which it is derived  ---
apply more generally to large classes 
of random matrices~\cite{Mehta},
only one example of which we have focused on here.

Finally, before concluding this section, it is important to realize that we have only partially
tackled the central problem that we face when discussing our random dark sector.
Thus far, we have focused on the extent to which a collection of $n_{\rm mat}$ matrices, each yielding
$N$ eigenvalues, together produce a set of $N n_{\rm mat}$ eigenvalues which match our expected ``emergent'' 
eigenvalue distributions.   This was done holding $N n_{\rm mat}$ fixed, so that larger values of $N$ required
fewer random matrices.        

This is fine for a mathematical study of random matrices.
However, for practical purposes, the physics question we wish to address is somewhat different.  
Given the fact that we have only one observable universe, we expect to see only one possible mass spectrum
for our DDM ensemble.
In other words, although the underlying symmetry-breaking process is random, in reality
we expect to get only one ``roll of the dice''.
Thus, the real question we need to face has to do with the extent to which a given {\it single}\/ roll of the dice 
--- \ie, the extent to which a {\it single}\/ randomly-generated $SU(N)$ matrix --- generates $N$ eigenvalues matching 
the expected eigenvalue-distribution function $n_{\rm SU(N)}$.   This is similar to the mathematical 
question we have already studied except that we now wish to hold $n_{\rm mat}=1$.

To study this question for any $N$, we randomly generate an $SU(N)$ matrix following the 
procedures outlined above, and calculate its $N$ eigenvalues.
We then place these eigenvalues into bins of equal size based on their magnitudes,
and calculate the goodness-of-fit  $\chi^2$ statistic 
\begin{equation}
        \chi^2 ~\equiv~  \sum_{i=1}^{n_{\rm bins}}  \frac{ (X_i - E_i)^2 }{E_i}
\label{chi2def}
\end{equation}
where $i$ labels the bin, where $n_{\rm bins}$ indicates the total number of bins,
where $X_i$ is the ``observed'' number of eigenvalues in the bin,
 and where $E_i$ is the ``expected'' number of eigenvalues in the bin according to the $n_{\rm SU(N)}$ 
eigenvalue-distribution function.
Given that we do not have an analytic form for $n_{\rm SU(N)}$, we obtain our values of $E_i$
via the methods outlined above, choosing a value of $n_{\rm mat}$ for each $N$ such
that a suitably large total number $n_{\rm eigs}$ of random eigenvalues are generated.  

The value of $\chi^2$ in Eq.~(\ref{chi2def}) represents the extent to which the eigenvalues
of a single randomly-generated $SU(N)$ matrix match $n_{\rm SU(N)}$, as desired.   Of course,
depending on the particular matrix generated, the corresponding $\chi^2$ could have values
which are extremely large or extremely small.   What interests us is therefore $\langle \chi^2\rangle$,
\ie, the {\it average}\/ value that $\chi^2$ might have
if we repeat this process $N'$ times.
To help us interpret the result, we then calculate a
Gaussian-equivalent significance $\sigma$ 
by comparing $\langle \chi^2\rangle$  to a $\chi^2$-distribution with $n_{\rm bins} - 1$ degrees of freedom in 
order to obtain a $p$-value, and then asking to what statistical significance $\sigma$ this $p$-value would 
correspond for a Gaussian distribution.  
This value of $\sigma$ thus  represents the degree to which, on average, the eigenvalues from a single random 
matrix seem {\it not}\/ to have been drawn from the expected  $n_{\rm SU(N)}$ 
distribution.  A lower significance $\sigma$ thus indicates better agreement with our expectations.

\begin{figure}[t]
  \includegraphics[width=0.35\textwidth, keepaspectratio]{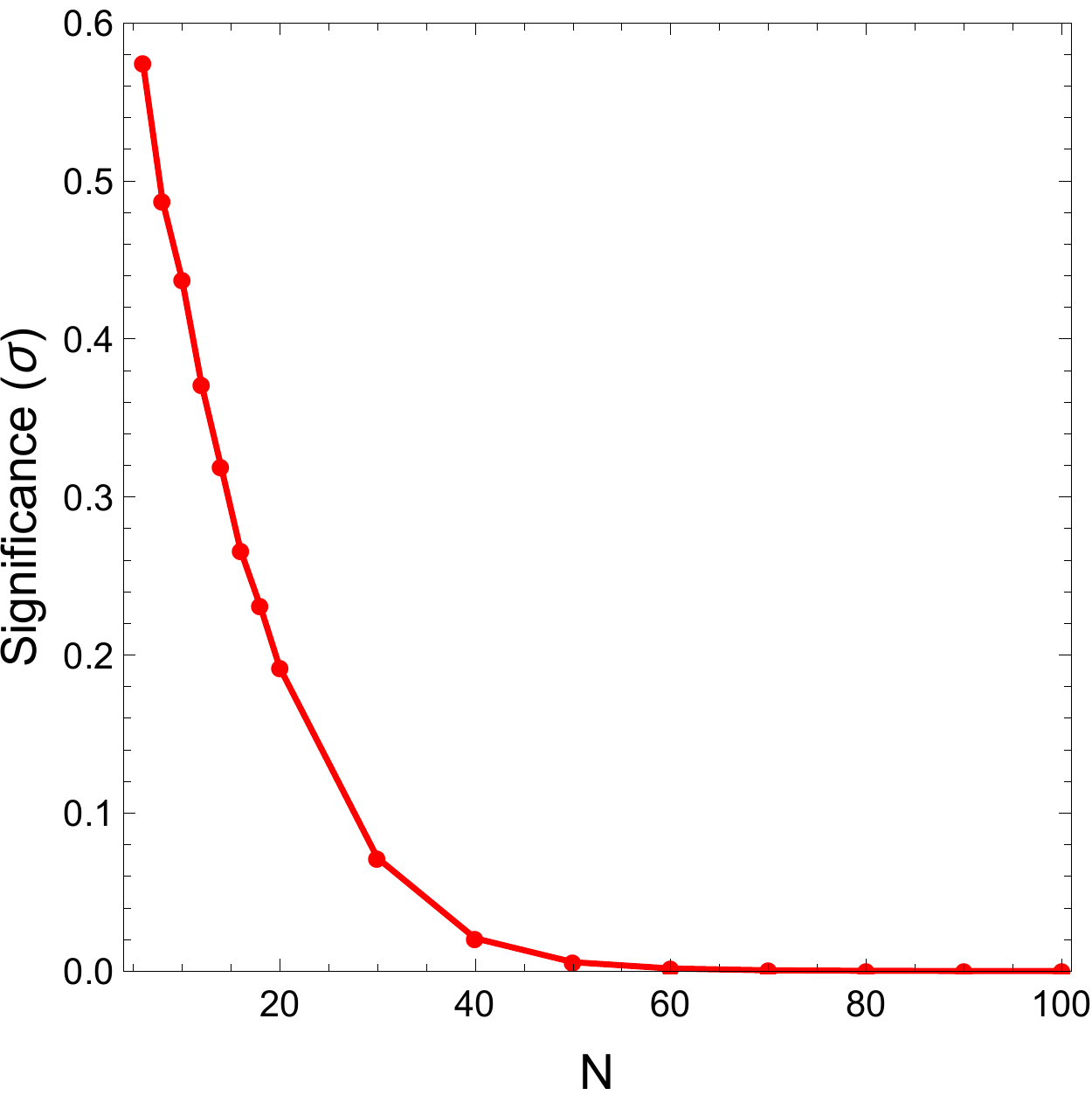}
  \caption{The accuracy with which the $N$ eigenvalues of a {\it single}\ randomly-chosen $SU(N)$ matrix
    match, on average, the $SU(N)$ probability distribution function
     $n_{\rm SU(N)}$, plotted as a function of $N$.
      As discussed in the text, this accuracy is quoted in terms of the traditional Gaussian-equivalent 
        significance $\sigma$ of the negative assertion
   that on average the eigenvalues derived from a single random matrix seem {\it not}\/ to be drawn 
from the expected distribution $n_{\rm SU(N)}$.
     Thus lower significance $\sigma$ indicates better agreement with $n_{\rm SU(N)}$.}
\label{fig:oneroll}
\end{figure}

Our results are shown in Fig.~\ref{fig:oneroll} for
the case of $n_{\rm eigs}=1.8\times 10^6$,
$N' = 10^5$, and $n_{\rm bins}=40$.
We have also taken $\tilde M=0$.
Note that none of the qualitative results in this figure would change significantly if other values were chosen.
We also remark that
when establishing our bins, we have 
we have assumed an eigenvalue range from $\tilde{m} = 0$ to $\tilde{m} = \tilde{m}_{\rm WS}^{\rm (max)}$, 
where  $\tilde{m}_{\rm WS}^{\rm (max)}$ is the maximum possible eigenvalue $\tilde{m}$ that can be obtained for the Wigner semicircle distribution for that same value of $N$.  
Although eigenvalues values larger than $\tilde{m}_{\rm WS}^{\rm (max)}$ are possible for the true $SU(N)$ ensemble,
we have already seen that such eigenvalues are exceedingly rare
and populate a low-statistics regime in which the ``expected'' bin population is extremely small.  
Including data in this regime therefore tends to skew the statistics in non-meaningful ways.   

Several features are immediately clear from Fig.~\ref{fig:oneroll}.   As expected,
we see that $\sigma$ decreases with increasing $N$, indicating that our $n_{\rm SU(N)}$ distributions ---
oscillations and all ---  
become increasingly precise as $N$ grows large.
In fact, given the results in Fig.~\ref{fig:oneroll},
we find that 
\begin{equation}
       \sigma ~\approx ~ (1.83)\, e^{- (0.114)\,N}  ~~~~~   {\rm for}~~N\gg 1~. 
\end{equation}
We thus see that the accuracy of our emergent mass distribution $n_{\rm SU(N)}$
actually grows {\it exponentially}\/ with $N$.
But perhaps most importantly, we see that our $\sigma$ values are themselves
extremely small {\it throughout}\/ the range of $N$ plotted, even for $N$  
which are relatively small.   Thus we can conclude that our emergent eigenvalue distributions $n_{\rm SU(N)}$ 
do an extremely good job of describing the distributions of masses across
the dark sector ---
even if the symmetries governing the dark sector are not overwhelmingly large.


\section{Towards a DDM Ensemble:   
Couplings, Decay Widths, and Cosmological Abundances\label{sec:DetectionAbundance}}


In the previous section, we analyzed the emergent mass spectrum 
for the example model presented in Sect.~\ref{sec:Model} and showed that for $N\gg 1$ 
the density-of-states function $n_{\rm SU(N)}(\tilde{m})$ for the $\phi_i$
converges to the Wigner semicircle function $n_{\mathrm{WS}}(\tilde{m})$ given in Eq.~(\ref{eq:DensityOfStateswithm}).  
However, in order for these particles 
to constitute a viable DDM ensemble, they must collectively manifest an appropriate 
balancing between decay widths and abundances.  In this section, we discuss how
a spectrum of decay widths and abundances can be generated for the $\phi_i$ and 
how the interplay between the corresponding scaling relations and the density-of-states
function derived in the previous section can give rise to a DDM-like balancing 
between decay widths and abundances across the ensemble as a whole.  
Indeed, as we shall see,
a density-of-states function which decreases with the mass $m$ 
 --- as emerges for our random ensemble ---
can accommodate a broader range
of scaling relations between decay width and mass, and between abundance and mass,
than is possible for other distributions.  

Note that here and in the following, indices $i,j=1,2,...,N$ now label 
the different mass eigenstates $\phi_i$.

\subsection{Dark-Matter Decay\label{sec:Decay}}

We begin with a discussion of the decay properties of the ensemble constituents. 
The toy model introduced in Sect.~\ref{sec:Model} involves only two additional
matter fields beyond those of the SM: the $SU(N)$ multiplets $\phi$ and $\eta$.  Given
the symmetry structure of the model, these fields couple to the fields of the SM only via 
gravitational interactions.  Moreover, the $\IZ_2^\phi$ symmetry of the model forbids 
all scattering and decay processes wherein number of $\phi_i$ particles in the 
initial and final states differ by an odd number.  As a consequence, the lightest of the 
$\phi_i$ is absolutely stable.  Furthermore, the decay widths of the heavier ensemble
constituents are typically dominated by {\it intra-ensemble}\/ decays --- \ie, decays to 
final states involving other, lighter $\phi_i$.  The most relevant intra-ensemble decay
processes include $\phi_i \rightarrow \phi_j \eta_{a}$, 
$\phi_i \rightarrow \phi_j \phi_k \phi_\ell$, and, in cases in which the $SU(N)$
symmetry is local, $\phi_i \rightarrow \phi_j G^\mu_{a}$, where $G^\mu_a$ denotes one 
of the $SU(N)$ gauge fields.

While channels of this sort provide a natural decay mechanism for the ensemble 
constituents within the context of this minimal statistical DDM model, these channels
are phenomenologically problematic for two reasons.  First, the corresponding decay 
widths are typically far too large.    
Second, because the final states consist solely of hidden-sector particles, these decays 
are essentially invisible and yield no observable indirect detection signals.
Thus, in order to ensure phenomenological viability, we must find a way to suppress 
these contributions while at the same time arranging additional mechanisms through which 
the $\phi_i$ can decay.    

The simplest and most natural way of suppressing processes involving $\eta_a$ or $G^\mu_a$
in the final state is to arrange for the masses of the $\phi_i$ to be parametrically small 
in comparison with the masses of the $\eta_a$ and $G^\mu_a$.  
Indeed, this is not difficult to arrange.  The overall mass scale for the $\phi_i$ 
is determined by the mass parameter $M$ in Eq.~(\ref{eq:V0}) and by the parameter 
combination $\sqrt{\xi_1} v$.  By contrast, the overall mass scales for the $\eta_a$ and 
the $G^\mu_a$ are both independent of $M$ and $\xi_1$.  Thus, suppressing decays such
as $\phi_i \rightarrow \phi_j \eta_{a}$ and $\phi_i \rightarrow \phi_j G^\mu_{a}$  
is simply a matter of taking $M$ and $\xi_1$ to be sufficiently small. 
Indeed,  this parametric separation is 
analogous to the situation in the 
SM lepton sector, where Majorana masses and Yukawa couplings of the light fermions play 
the roles of $M$ and $\xi_1$, respectively.

Intra-ensemble decay processes such as $\phi_i \rightarrow \phi_j \phi_k \phi_\ell$, 
in which one of the $\phi_i$ decays to a final state involving only other, lighter ensemble 
constituents, can also naturally be suppressed in a number of ways.  First of all,
contributions to such processes which arise due to the quartic interaction in 
Eq.~(\ref{eq:V0}) can be suppressed simply by taking the coupling coefficient $\xi_\phi$,
which plays no other important role in the phenomenology of the model, to be sufficiently small.  
Moreover, the remaining contributions, which proceed via diagrams involving virtual $\eta_a$ or 
$G^\mu_a$ fields, can likewise be suppressed by taking $M$ and $\xi_1$ to be sufficiently small.   
Alternatively, intra-ensemble decay processes of this sort can also be forbidden simply by 
kinematical considerations.  For example, processes of the form 
$\phi_i \rightarrow \phi_j \phi_k \phi_\ell$ are kinematically forbidden unless  
$m_{\phi_i} > 3m_0$, where $m_0$ is the mass of the lightest ensemble
constituent.  If the mass spectrum of the theory is such that this condition is not
satisfied for even the heaviest of the $\phi_i$, such intra-ensemble decays cannot occur.
This is not difficult to arrange.  Indeed, $m_0$ is essentially determined by $M$, while 
the splitting between the masses of the heaviest and lightest of the ensemble constituents 
is essentially determined by $\sqrt{\xi_1} v$; thus, the lightest mass and the largest mass 
splitting are parametrically independent.  

Having discussed how contributions from processes such as 
$\phi_i \rightarrow \phi_j \phi_k \phi_\ell$, $\phi_i \rightarrow \phi_j \eta_{a}$,
and $\phi_i \rightarrow \phi_j G^\mu_{a}$ to the decay widths of the $\phi_i$ can be
suppressed, we now turn to examine how decay widths of the correct order of magnitude can be 
generated for these fields.
For concreteness, we present an example of how the ensemble constituents 
can be coupled to the fields of the SM --- and in particular the photon field.  We 
emphasize that this is merely one example, and that other self-consistent 
coupling scenarios which lead to alternative decay phenomenologies for the $\phi_i$
are possible as well.  

As discussed above, our first step is to break the $\IZ_2^\phi$ symmetry.
In order to do this in a controlled 
manner, we introduce an additional field $B$ which is odd under $\IZ_2^\phi$ and 
transforms in the fundamental representation of $SU(N)$.
Given the symmetry structure of the visible and hidden 
sectors, the simplest method of coupling the $\phi_i$ to the photon field is via
an effective operator of the form
\begin{equation}
  \mathcal{O}_D  ~=~ \frac{c}{\Lambda^2} ( \phi^\dagger B + B^\dagger \phi ) 
    F^{\mu \nu} F_{\mu \nu}~,
  \label{eq:DecayOperator}
\end{equation}
where $F^{\mu\nu}$ is the field-strength tensor for the SM photon field, 
where $\Lambda$ is the cutoff scale of the effective theory, and where $c$ is a dimensionless
operator coefficient.  

We shall assume that $B$ acquires a VEV due to some additional 
dynamics.  Provided that this dynamics is uncorrelated with the dynamics which gives 
rise to the mass matrix for the $\phi_i$, the direction of this VEV in field space is 
arbitrary.  For convenience, we choose to parametrize the VEVs of the individual components 
$B_i$ of $B$ (in the mass eigenbasis of the $\phi_i$) as
\begin{equation}
  \langle B_i \rangle ~\equiv~ b \beta_i~.
\end{equation}  
Here, the $N$ different $\beta_i$ are dimensionless and subject to the constraint
\begin{equation}
  \sum_i \beta_i^* \beta_i ~=~ 1~.
  \label{eq:BetaSqRel}
\end{equation}
Note that since the $\beta_i$ characterize the direction of the VEV in field space,
all of them are generically non-zero.  By contrast, $b$ is a parameter with dimensions 
of mass which characterizes the overall magnitude of the collective contribution to 
the breaking of 
$\IZ_2^\phi$ and $SU(N)$ from the $\langle B_i\rangle$.
Since the results derived in Sect.~\ref{sec:RandomMatrixGeneration} are predicated on
the $\langle \eta_a \rangle$ being the dominant source of $SU(N)$ symmetry-breaking,
we require that $b \ll v$.  Moreover,  as discussed in greater detail below, $b$ controls 
the strength of the effective couplings between the ensemble constituents and the photon 
field, and therefore the lifetimes of these constituents.  Thus, phenomenological 
considerations associated with dark-matter decay --- including, at the very least, the 
requirement that the longest-lived particle in the ensemble have a lifetime that exceeds
the present age of the universe --- likewise constrain $b$ to be quite small.

Expanding the $B_i$ around their VEVs in Eq.~(\ref{eq:DecayOperator}) yields an
effective tree-level interaction between each of the ensemble constituents and a
pair of photons:  
\begin{equation}
  \mathcal{O}_D ~=~\frac{b}{\Lambda^2} F^{\mu\nu} F_{\mu\nu}
    \sum_i \big( \phi_i^\dagger \beta_i + \beta_i^* \phi_i\big) + \ldots
  \label{eq:DecayOperatorVEVed}
\end{equation}
Note that in principle, the $\langle B_i \rangle$ generically breaks any remaining 
continuous symmetries under which the $\phi$ are charged, including the residual
$U(1)$ symmetry associated with phase rotations of each complex scalar $\phi_i$.
As a result, the $\langle B_i \rangle$ generically break the degeneracy between the 
real and imaginary components of each $\phi_i$.  Thus, in principle, these degrees of
freedom should be treated as distinct real fields.  However, because
the induced mass splitting between these fields is proportional to $b$, which is 
constrained to be small in comparison with the masses of the $\phi_i$, as discussed
above, this splitting has a negligible effect on the mass spectrum of the theory.  
Therefore, in practice, we may safely neglect the splitting between the real and 
imaginary parts of the $\phi_i$ and continue to treat the ensemble constituents as
complex fields in what follows.

Under the assumption that the operator in Eq.~(\ref{eq:DecayOperatorVEVed}) provides
the leading contribution to the decay width $\Gamma_i$ of each ensemble constituent 
$\phi_i$, the decay rate of each component is then given by
\begin{equation}
  \Gamma_i ~ \approx ~ \frac{b^2 m^3_{i}}{2\pi \Lambda^4} |\beta_i|^2~. 
\label{eq:DecayRate}
\end{equation}
It is noteworthy that the effect of the randomness in these decay widths is reflected
in the coefficients $|\beta_i|^2$ which in turn 
must satisfy the constraint in Eq.~(\ref{eq:BetaSqRel}).
This indicates that while randomness can produce ${\cal O}(1)$ fluctuations
in the sizes of these decay widths, these decay widths must nevertheless satisfy 
the bound
\begin{equation}
  \Gamma_i ~ \leq ~ \frac{b^2 m^3_{i}}{2\pi \Lambda^4} ~. 
\label{eq:DecayRate2}
\end{equation}
These decay widths therefore cannot grow without limit.
As a result, 
by adjusting the parameters in Eq.~(\ref{eq:DecayRate2}), 
it is possible to ensure that our ensemble does not violate observational limits on dark-matter
decays to photons (see, \eg, Ref.~\cite{Slatyer:2012yq})  ---
all despite the inherent randomness introduced into
the decay widths through the coefficients $\beta_i$.

\subsection{Annihilation and Abundances}

A number of mechanisms exist through which a spectrum of primordial abundances for a
set of dark-sector particles can be established.  For concreteness, we focus here on 
one particular mechanism, namely a variation~\cite{DesigningDDMPaper} of thermal freeze-out 
which yields particularly interesting scaling relations between the masses and abundances of 
ensemble constituents.

The spectrum of primordial abundances $\Omega_i$ for the $\phi_i$ depends principally 
on the masses of these particles and on the cross-sections for the processes through which 
they annihilate.  One of the simplest ways of arranging a set of cross-sections of
the correct order of magnitude is to introduce~\cite{DesigningDDMPaper} 
an additional dark fermion $\psi$ with a mass 
$m_\psi < m_0$ into which the $\phi_i$ can annihilate (where $m_0$ is the mass of the lightest
ensemble constituent), as well as an additional Abelian 
gauge group $U(1)_\chi$ under which both $\psi$ and the multiplet $\phi$ are charged.
For phenomenological reasons, we shall also assume that $U(1)_\chi$ is broken by some 
additional dynamics, and therefore that the $U(1)$ gauge field $\chi^\mu$ is massive.  In what follows,
we shall assume that the breaking of this symmetry is essentially unrelated to the
breaking of $SU(N)$.  In other words, we shall assume that the Goldstone boson 
which provides the longitudinal polarization of $\chi^\mu$ transforms as a singlet 
under $SU(N)$.  Under this assumption, 
all components of $\phi$ will have identical $U(1)_\chi$ charges 
and the couplings between $\chi^\mu$ and the mass eigenstates 
$\phi_i$ may be taken to be diagonal.

In this scenario, the cross-section for dark-matter annihilation 
processes of the form $\phi_i^* \phi_i \rightarrow \overline{\psi} \psi$ receives a
contribution from diagrams involving an $s$-channel $\chi^\mu$, whereas the cross-section
for coannihilation processes of the form $\phi_i^* \phi_j \rightarrow \overline{\psi} \psi$ 
with $i\not= j$
receives no such contribution.  Contributions to the cross-sections for annihilation
process of the form $\phi_i^* \phi_i \rightarrow \phi_j^* \phi_j$ with $m_i > m_j$ (also
involving an $s$-channel $\chi^\mu$) are also generated.  Moreover, contributions to the 
cross-sections for processes of the form $\phi_i^* \phi_i \rightarrow \chi^\mu \chi_\mu$ 
are generated for ensemble constituents with $m_i > m_\chi$ directly from the 
corresponding gauge-kinetic terms in the field Lagrangian.  However, in situations in which 
the effective coupling $g_\psi$ between $\chi^\mu$ and the $\psi$ field is much larger
than the coupling $g_i$ between $\chi^\mu$ and $\phi_i$, the
$\phi_i^* \phi_i \rightarrow \overline{\psi} \psi$ process will dominate. 
We shall henceforth assume that $g_\psi \gg g_i$ for all $\phi_i$, and thus that 
the $\phi_i^* \phi_i \rightarrow \overline{\psi} \psi$ process provides the dominant 
contribution to the annihilation cross-section of all of the $\phi_i$.
Moreover, we shall also take 
$g_i^2 N \ll 1$ and $g_i^2 N \ll g_\psi^2$ in order to ensure
that the annihilation process remains perturbative at all relevant scales.  

Under these assumptions, the thermally averaged annihilation cross-section 
$\langle \sigma_A v \rangle_i$ for each ensemble constituent is entirely determined 
by the masses $m_i$, $m_\psi$, $m_\chi$ and the couplings $g_i$, $g_\psi$.
Following the analysis of Ref.~\cite{DesigningDDMPaper}, we find that the 
corresponding thermal contribution to the relic abundance of each ensemble 
constituent takes the form 
\begin{equation}
  \Omega_i ~\propto~ \langle \sigma_A v \rangle_i^{-1} 
    ~\propto~ \frac{m_i^2}{g_i^2 g_\psi^2 }
    \left(1 - \frac{m_\chi^2}{4m_i^2} \right)^2~.
  \label{eq:Abundance}
\end{equation}
Alternatively, the $\Omega_i$ may be expressed in terms of the mass $m_0$, abundance 
$\Omega_0$, and $U(1)_\chi$ coupling $g_0$ of the lightest ensemble constituent.
The result is
\begin{equation}
  \Omega_i ~=~ \Omega_0 \left(\frac{m_0g_0}{m_ig_i}\right)^2 
    \left(\frac{4m_i^2 - m_\chi^2}{4m_0^2 - m_\chi^2}\right)^2~.
  \label{eq:AbundanceInTermsOfOmega0}
\end{equation}
We emphasize that in this expression the subscript `0' is merely shorthand 
for that value of $i=1,...,N$ for which $m_i$ is minimized.

In interpreting the result in Eq.~(\ref{eq:AbundanceInTermsOfOmega0}), 
it is instructive to consider the case in which the
couplings between $\chi^\mu$ and all of the $\phi_i$ are identical --- \ie, 
the case in which $g_i = g_0$ for all $\phi_i$.
Thus, as discussed in Ref.~\cite{DesigningDDMPaper}, we see that in the limit in which 
$m_i \gg m_\chi$, the abundance $\Omega_i$ increases with the mass of the ensemble 
constituent.  In particular, $\Omega_i \propto m_i^2$.  By contrast, in the opposite
limit in which $m_i \ll m_\chi$, we see that $\Omega_i \propto m_i^{-2}$.  Thus, in
this limit, the heavier ensemble constituents have smaller abundances.  Finally, we
note that in the special case in which $2m_i \approx m_\chi$ and annihilation of a 
particular ensemble constituent occurs on resonance, we obtain $\Omega_i \propto \Gamma_\chi^2 $, 
where $\Gamma_\chi$ is the total decay width of $\chi$.

\subsection{Random Ensembles as DDM Ensembles\label{sec:DDMatlast}}

We now seek to determine whether the scaling relations we have found for our random
ensemble are
consistent with the phenomenological balancing requirements that are
the cornerstone of the DDM framework.
As discussed in Refs.~\cite{DDM1,DDM2}, these balancing requirements may be expressed as follows.
First, we use our relation between decay widths $\Gamma$ and masses $m$ in order
to express our abundance function $\Omega(m)$ 
and density-of-states function $n(m)$ 
in terms of $\Gamma$ rather than $m$.
Note that while  $\Omega(m)$ and $n(m)$ are each 
densities {\it per unit mass},
what we now seek are 
the corresponding densities $\Omega(\Gamma)$ and $n(\Gamma)$ 
 {\it per unit decay width}:
\begin{equation}
  \Omega(\Gamma) ~=~ \Omega(m) \left|\frac{dm}{d\Gamma}\right|~,~~~
  n(\Gamma) ~=~ n(m) \left|\frac{dm}{d\Gamma}\right|~.
\end{equation}
Given these new functions, 
it was then shown in Ref.~\cite{DDM1} that if 
$\Omega(\Gamma)$ scales with $\Gamma$ according to 
$\Omega(\Gamma) \propto \Gamma^\alpha$ for some scaling coefficient $\alpha$,
and if  $n(\Gamma) \propto \Gamma^\beta$ for some scaling coefficient $\beta$,
then the requirement
\begin{equation}
         x ~\equiv~ \alpha + \beta ~\lsim ~ -1~
\label{criterion}
\end{equation}
serves as a good  rudimentary criterion for assessing
whether decay widths are balanced against abundances in the appropriate manner to yield
a viable DDM ensemble.

For the random toy model we have presented here, the abundance function $\Omega(m)$ and 
density-of-states function $n(m)$, expressed as functions of the constituent mass $m$, 
are given in Eqs.~(\ref{eq:AbundanceInTermsOfOmega0}) and~(\ref{eq:DensityOfStateswithm}) for
$N\gg 1$, respectively.  
Likewise, the result in Eq.~(\ref{eq:DecayRate}) implies that 
$m \propto \Gamma^{1/3}$, whereupon we see that $|dm/d\Gamma| = \Gamma^{-2/3}$.
As discussed above, $\Omega_i \propto m_i^{-2}$
in the $m_i \ll m_\chi$ limit; 
thus in this limit we have $\Omega(m) \propto \Gamma^{-2/3}$
and $\Omega(\Gamma)\propto \Gamma^{-4/3}$.
We therefore find $\alpha = -4/3$ 
for this toy model.  By contrast, the density-of-states
function for the Wigner semicircle distribution which emerges for large $N$
does not scale with $m$ according to a power law $n(m) \propto \Gamma^\beta$ 
with a constant index $\beta$, but rather according to a relation
$n(m) \propto \Gamma^{\delta(m)}$ with a variable index $\delta(m)$.
Since $n(m)$ decreases monotonically with $m$, this variable index is bounded from
above:   $\delta(m) \leq 0$.  
This implies a corresponding bound 
$\beta(\Gamma) \leq -2/3$ on the variable index $\beta(\Gamma)$.  
Our random toy model therefore has
\begin{equation}
  x(\Gamma) ~\leq~ -2~,
\end{equation}
and this satisfies Eq.~(\ref{criterion}) for all values of $\Gamma$.  We thus conclude that our random
ensemble indeed displays an appropriate balancing between lifetimes and abundances, as needed in order to serve as
a potentially viable DDM ensemble.

It is important to note that this success is not merely 
restricted to situations in which our decay widths are dominated by dimension-five operators,
such as we have considered above.
In general, if $d$ is the dimension of the dominant decay operator
and the daughter particles are light,
dimensional analysis implies that 
$\Gamma \propto m^{2d-7} / \Lambda^{2d-8}$.
Repeating the above steps assuming $\Omega\propto m^{-2}$ 
then yields $\alpha = (6-2d)/(2d-7)$ and $\beta \leq (8-2d)/(2d-7)$,
whereupon we see that $x=\alpha+\beta \leq -2$, exactly as before.
Thus the suitability of our random ensemble to serve as a DDM ensemble
is a fairly robust phenomenon.

A few additional comments are in order.  First of all, while we have taken the field 
$\psi$ into which the ensemble constituents annihilate to be a dark-sector fermion
for concreteness, there is nothing special about this choice.  For example, 
essentially none of our results would change were we to have taken $\psi$ to be a 
scalar.  Indeed, the result in Eq.~(\ref{eq:AbundanceInTermsOfOmega0}) does not 
depend on the spin of $\psi$.  Since the coupling $g_\psi$ is independent 
of $m_\chi$, the scaling of $\langle \sigma_A^i\rangle$ with $m_i$ is determined 
by dimensional analysis alone.  Moreover, $\psi$ need not be a dark-sector particle
at all.  For example, a SM fermion could potentially play the role of $\psi$, 
provided that phenomenological constraints from indirect detection, collider physics, 
\etc, are satisfied.    

Finally, we observe that it is possible to get a sense of the overall mass scale for
the DDM ensemble constituents for which a thermal relic density can be successfully 
obtained by analogy with the case of a standard WIMP.
Once again, we focus our attention on the perturbative regime and take $m_i \ll m_\chi$
for all $\phi_i$, so that $\Omega_i \propto m_i^{-2}$.  In the limit in which 
$M \gg \sqrt{\xi_1}$ and the masses of the $\phi_i$ are approximately degenerate ---
\ie, $m_i \approx m_0$ for all $\phi_i$ --- we find that the total abundance of the 
ensemble takes the form
\begin{equation}
  \Omega_{\mathrm{tot}} ~\propto~ \frac{N m_\chi^4}{g_0^2 g_\psi^2 m_0^2}~.
\end{equation}  
For a standard WIMP which receives its abundance via thermal freeze-out, the usual 
constraint that arises due to the requirement that the annihilation cross-section be 
perturbative is $m_{\mathrm{WIMP}} \lesssim 300~\tev$~\cite{GriestMassLimit}.
The corresponding constraint on a DDM ensemble of this sort in the degenerate limit
in which the $m_i$ are approximately equal becomes
\begin{equation}
  m_i ~\lesssim ~ \frac{300~\tev}{N}~.
\end{equation}
 

\section{Indirect-Detection Signals\label{sec:IndirectDet}}


We now discuss one possible observational signal which arises in this DDM scenario.
In particular, we examine the spectrum of high-energy photons produced by
the annihilation and decay of the $\phi_i$.
In principle, contributions to the photon spectrum can arise in this model both from
the dark-matter decay process $\phi_i \rightarrow \gamma \gamma$ and from the 
annihilation process $\phi_i^* \phi_i \rightarrow \bar{\psi} \psi$, followed by the
decay of $\bar{\psi}$ and $\psi$ and into SM particles. 
For simplicity, because the annnihilation contribution depends on additional model 
inputs not directly related to the decay widths or abundances of the $\phi_i$ --- and, 
in particular, on the decay properties of $\psi$ --- we concentrate on the decay 
contribution.  
 
Under the assumption that the operator in Eq.~(\ref{eq:DecayOperatorVEVed}) 
dominates the width of each of the $\phi_i$, the primary injection 
spectrum of photons from dark-matter decay --- \ie, the differential photon flux 
per unit energy $E_\gamma$, including the individual contributions 
from all ensemble constituents --- takes the form
\begin{equation}
  \frac{dN}{dE_\gamma} ~\propto~ \sum_i 
    \frac{\Omega_i \Gamma_i}{m_i} \delta(m_i - 2E_\gamma)~.
\label{abund}
\end{equation}
The overall normalization of this 
injection spectrum depends on the astrophysical properties of the object(s) 
under observation.  Interesting possibilities might include, \eg, the galactic center 
and the halos of Milky-Way dwarf galaxies.  However, we can assume that the 
local energy densities 
of the $\phi_i$ within the object(s)
under study are proportional to the corresponding cosmological energy densities $\rho_i$.
In this case, the {\it shape}\/ of the injection spectrum depends on particle-physics
considerations alone.   

\begin{figure*}[t]
  \includegraphics[width=0.3\textwidth, keepaspectratio]{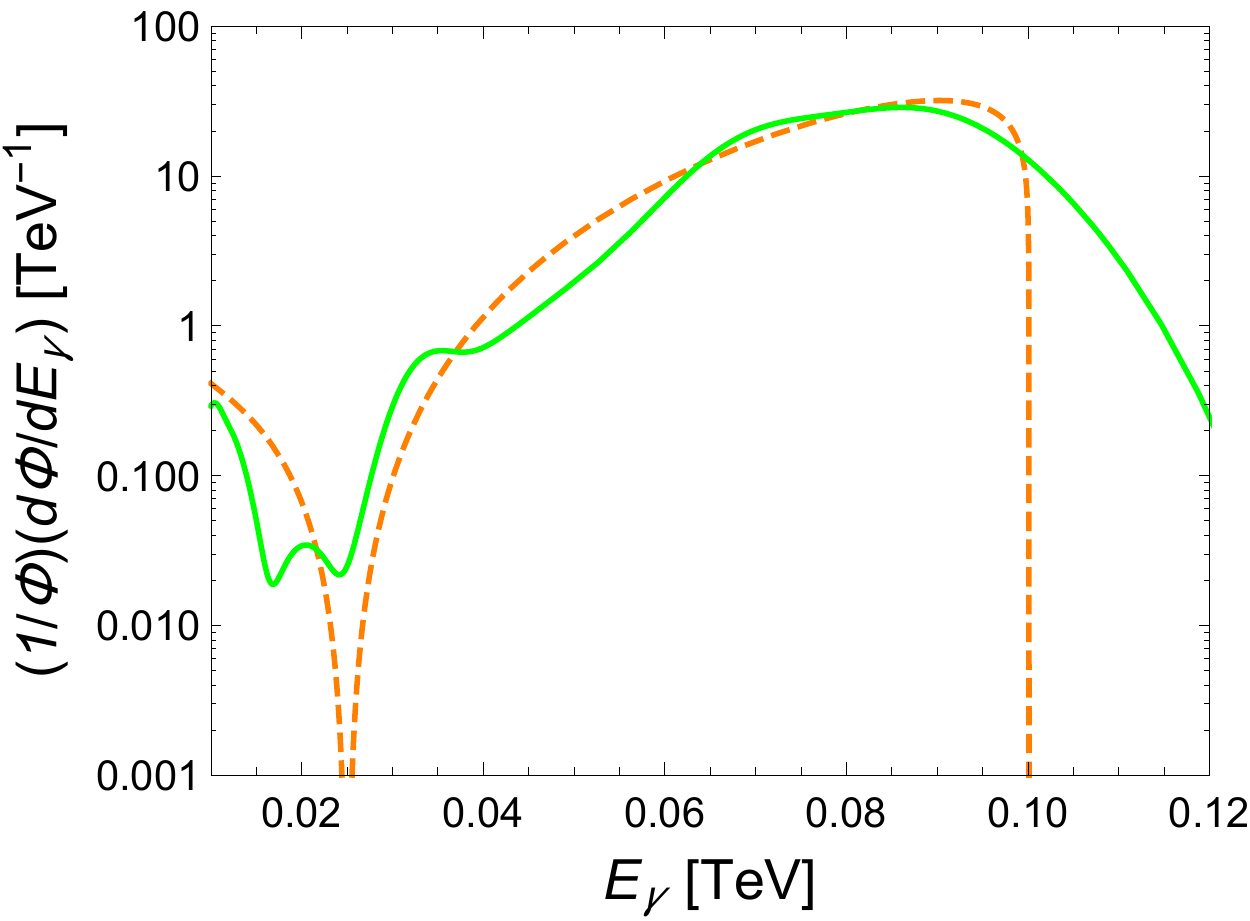}~~~
  \includegraphics[width=0.3\textwidth, keepaspectratio]{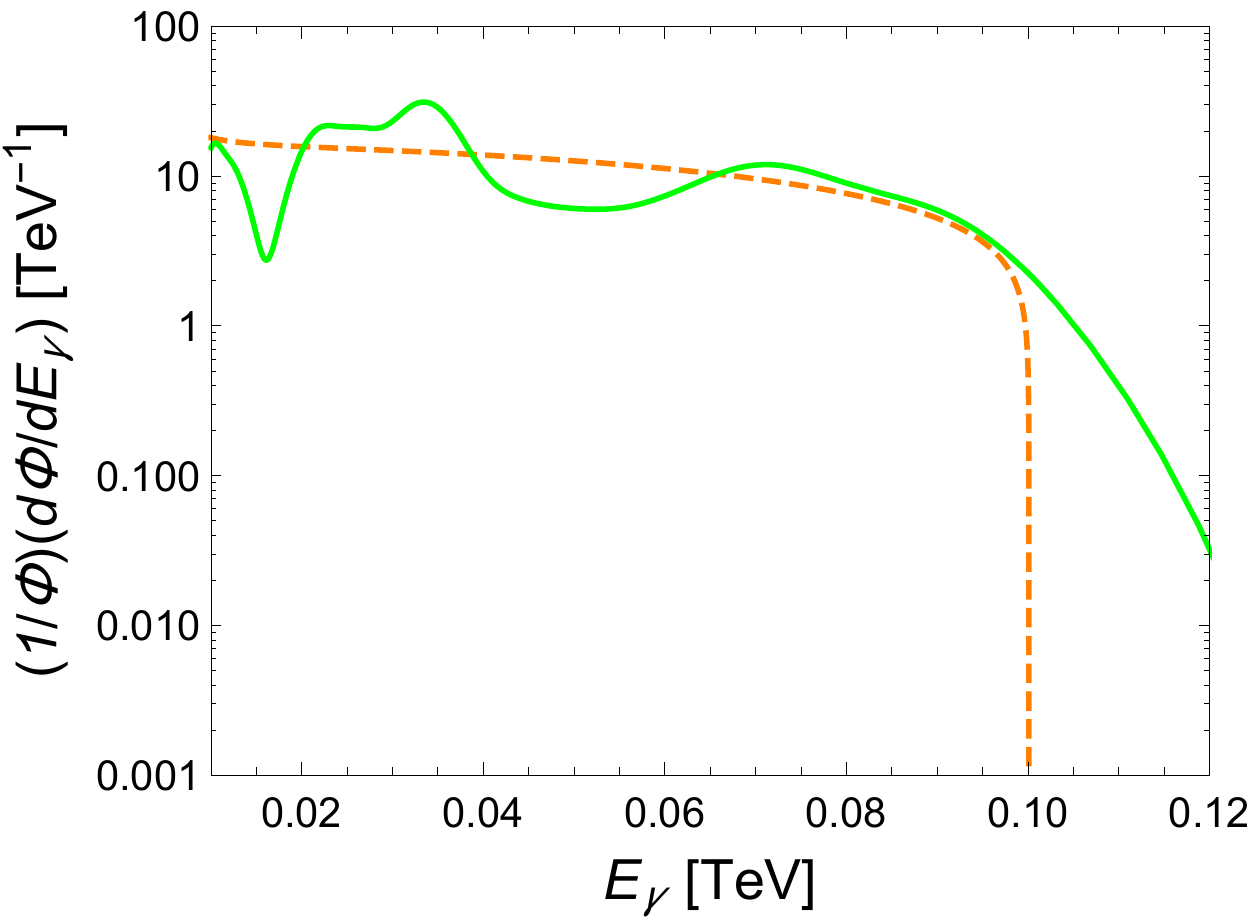}~~~
  \includegraphics[width=0.3\textwidth, keepaspectratio]{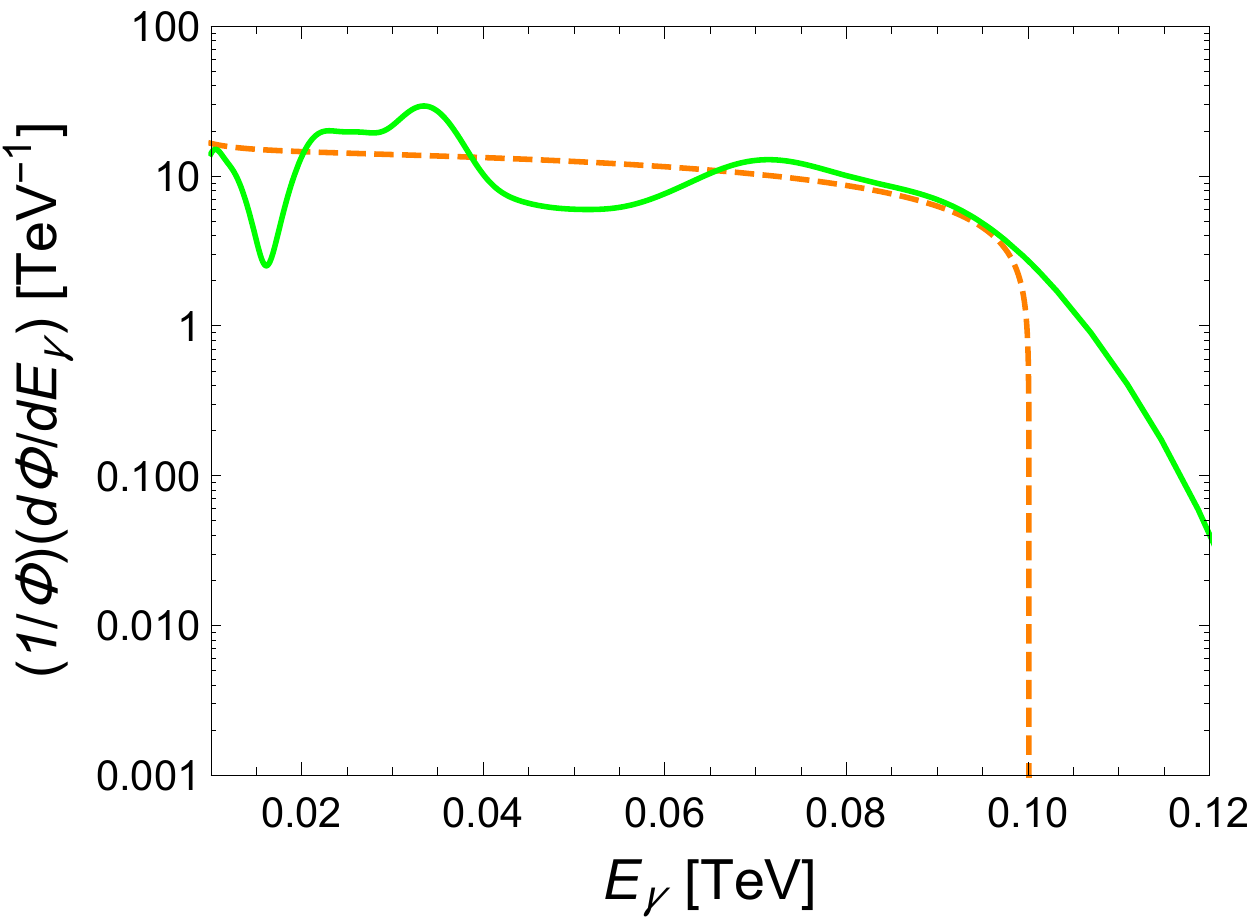}
  \caption{The normalized photon spectra associated with a set of benchmark DDM models
     which experience decays through the operator in Eq.~(\ref{eq:DecayOperatorVEVed}).
     In each case, the DDM ensemble is characterized by the parameter choices 
     $N=100$, $M=10$~GeV, and $\sqrt{\xi_1 v^2} = \sqrt{2}$~TeV, while  the left, center, and 
     right panels show the results obtained for
     the cosmological abundances $\Omega_i$ 
     that would result from $U(1)_\chi$ gauge-boson masses 
     $m_\chi = 100$~GeV, $1$~TeV, and $10$~TeV, respectively.  
     Given the inherent randomness in the $\phi_i$ mass matrix and the coupling parameters 
     $\beta_i$ in Eq.~(\ref{eq:DecayOperatorVEVed}), the green curve appearing in 
     each panel represents one possible spectrum which might be observed at a gamma-ray 
     detector with a Gaussian energy resolution of $\Delta E_\gamma / E_\gamma = 0.09$.
    The green curves in all three panels correspond to the same choice of
    underlying random parameters.
      By contrast, the orange curve appearing in each panel represents the corresponding 
     ``baseline'' injection spectrum obtained by taking the analytic expression for $n_{\rm WS}(m)$ 
     as the density of states for the ensemble and by taking a universal coupling 
     $|\beta_i| = N^{-1/2}$ in Eq.~(\ref{eq:DecayOperatorVEVed}), while ignoring 
     detector effects.}
  \label{fig:DMDecaySpectrum}
\end{figure*}

We now illustrate the qualitative features which arise in the photon spectrum of 
scenarios of this sort.
In Fig.~\ref{fig:DMDecaySpectrum}, we display the contribution to the photon 
flux from $\phi_i$ decay in a set of benchmark DDM models with $N=100$, $M = 10$~GeV, and 
$\sqrt{\xi_1 v^2} = \sqrt{2}$~TeV.~  The left, center, and right panels
of Fig.~\ref{fig:DMDecaySpectrum} correspond to $U(1)_\chi$ gauge-boson masses of 
$m_\chi = 100$~GeV, $1$~TeV, and $10$~TeV, respectively, where these
masses enter into the calculation of the cosmological abundances $\Omega_i$ in
Eq.~(\ref{abund}).  
The green curve in each panel represents one possible spectrum of signal events 
which might be observed by a gamma-ray detector with a Gaussian energy 
resolution $\Delta E_\gamma / E_\gamma = 0.09$, given the inherent randomness 
in the $\phi_i$ mass matrix and the coupling parameters $\beta_i$.   
Note that we have normalized 
each spectrum shown such that the total area under the curve is unity.   
By contrast, the orange curve shown 
in each panel represents the corresponding injection spectrum obtained by taking 
the analytic expression for $n_{\mathrm{WS}}(m)$ as the density of states for
the ensemble, and by taking a universal coupling $\beta_i = N^{-1/2}$ between
each $\phi_i$ and the photon field, all while ignoring detector effects.  
Thus, the orange curve can be viewed as representing the overall reference ``baseline'' 
around which the corresponding actual injection spectrum fluctuates due 
to the inherent randomness in the masses and couplings.

The gamma-ray spectra shown in Fig.~\ref{fig:DMDecaySpectrum} display a 
number of distinctive features.  One such feature is the characteristic shape of the 
spectral baseline (\ie, the orange curve), which is primarily determined 
by the density-of-states function $n_{\mathrm{WS}}(n)$ for our statistical DDM ensemble.
Indeed, in the center and right panels of the figure, the influence of this 
density-of-states function on the spectral envelope is unmistakable.
In the left panel, which corresponds to the case in which $m_\chi = 100$~GeV, 
the shape of the envelope is modified at energies around $E_\gamma \approx 25$~GeV
due to the effect of the annihilation resonance, which suppresses the abundances
(and therefore the photon-flux contributions) of ensemble constituents with 
masses $m_i \sim m_\chi/2 \approx 50$~GeV.  ~Nevertheless, even in this case, the
influence of the density-of-states function on the spectral baseline is still  
evident --- especially at large $E_\gamma$.
   
Another distinctive feature which emerges in the gamma-ray spectra of statistical 
DDM ensembles is the pattern of fluctuations in the observed spectrum around the 
spectral baseline.
Indeed, in the regime in which the scale of the splitting between the $m_i$ exceeds the energy resolution
of the detector,
significant fluctuations are apparent across the relevant
range of $E_\gamma$.  
These fluctuations (along with
detector effects) can obscure features which would otherwise be apparent in the gamma-ray 
spectrum.  For example, the pronounced dip in the spectral baseline due to the 
annihilation resonance is only partially evident in the observed gamma-ray spectrum for the 
corresponding DDM ensemble.   
By contrast, we see that in the regime in which $E_\gamma$ is large and the 
energy resolution of the detector is less than or comparable to the rough scale of the splitting
between the 
mass eigenvalues $m_i$, the effect of the random alignment of the $\langle B_i\rangle$ is 
smeared out by detector effects.  
Consequently, in this regime, the observed spectrum is reasonably well approximated by the spectral baseline
up to the cutoff at $E_\gamma \approx 100$~GeV, which corresponds to the upper limit of the Wigner-Semicircle distribution for this choice of parameters.   The divergence between the observed spectrum and the spectral baseline 
above this cutoff is solely a result of detector smearing.

We note that the flux spectrum shown in left panel of 
Fig.~\ref{fig:DMDecaySpectrum} displays fluctuations with magnitudes as large as 
50\% or more of the overall baseline.  This is comparable to the fractional uncertainty 
in the measurement of the isotropic diffuse gamma-ray background at 
$E \sim 100$~GeV~\cite{Ackermann:2014usa}.  This implies that the potentially 
identifiable signals of statistical DDM ensembles of this sort may manifest themselves 
not only in the distinctive shape of the overall baseline associated with the gamma-ray spectrum, 
but also in the pattern of fluctuations in that spectrum.


\section{Conclusions\label{sec:Conclusion}}

In this paper, we studied the possibility that the properties of the dark sector
are dominated by processes which are essentially random.
We considered a dark sector composed of an ensemble 
of $N$ individual 
components
with differing masses, cosmological abundances, and couplings
to the SM,
and constructed a toy model in which the mass spectrum associated
with this ensemble is determined through an essentially random
breaking of an internal $SU(N)$ dark-sector symmetry.
Even though this mass spectrum is determined randomly,
we were nevertheless
able to bring the machinery of random-matrix theory to bear in order to
derive predictions for the probability distribution of mass eigenvalues
for these component fields.  In this way, we were able to obtain a statistical prediction for a
density-of-states function $n(m)$ 
for the ensemble as a whole --- a prediction 
which grows more and more robust as the number $N$ of ensemble constituents increases.  
Moreover, we found that this emergent density of states decreases as a function of mass and actually
has an upper limit $m_{\rm max}$ beyond which $n(m)=0$ --- behavior which is quite unlike
the density-of-states functions for all other DDM ensembles which have previously been 
discussed in the literature.

Given these results for the dark-sector mass spectrum,
we then proceeded to demonstrate that 
a set of corresponding cosmological abundances and decay widths for the 
ensemble constituents can be generated via well-established mechanisms --- thermal 
freeze-out and the controlled breaking of a stabilizing symmetry, respectively.
We then found that in scenarios of this sort, the fundamental scaling relations
which govern the ensemble satisfy the basic criteria to be interpreted as potentially
viable DDM ensembles --- ensembles in which cosmological abundances and decay widths satisfy
certain balancing relations across all constituents.
Thus, we were able to demonstrate that randomness in the dark sector coexists 
quite naturally with DDM.~
Finally, we explored one possible observational signature of random DDM ensembles
of this sort, namely the indirect detection of high-energy gamma rays produced 
from the decays of the ensemble constituents, and evaluated the prospects for the 
detection of such signals.  

There are many ways in which the analysis of this paper might be generalized and extended.
As we have discussed in the Introduction,
there are three aspects of a dark-sector ensemble which are critical 
in determining the resulting phenomenology:
the spectrum of constituent masses, the spectrum of constituent cosmological abundances,
and the spectrum of constituent decay widths into SM states.
While the first of these describes the properties of the ensemble unto itself,
the second and third depend upon further information concerning how the ensemble 
emerges within a particular cosmological history
and/or couples to SM states.
In this paper, we investigated the case in which only the first of these --- namely
the mass spectrum --- is determined randomly.
Indeed, as have seen, the simple random mechanism which gives rise to
the density-of-states function for our ensemble is essentially unrelated to the 
physics which determines the abundances and decay widths of the ensemble constituents.

Given this,
we may view our random mechanism as a fundamental ``kernel'' for establishing an
ensemble with a particular density-of-states function, a kernel which can be incorporated as
an ingredient in any number of alternative scenarios.  As discussed in 
Sect.~\ref{sec:RandomMatrixGeneration}, the fact that the density of states for such
ensembles decreases with increasing $m$ implies that this ``kernel'' should be compatible
with a wide variety of scenarios for abundance and decay-width generation.  Indeed,
a broad class of models could be developed around ensembles of this sort  
whose detailed phenomenological implications might exhibit interesting and unanticipated characteristics.
It would also be interesting to explore further scenarios in which the
mechanisms for generating cosmological abundances and SM decay widths  
are themselves based on random processes.

It would also be interesting to consider more broadly the methods 
by which randomness in the dark sector might be detected.
Identifying characteristic features in the gamma-ray 
spectra of those statistical DDM ensembles,
as we have discussed in Sect.~\ref{sec:IndirectDet}, 
may represent only one possible way of distinguishing
between such ensembles experimentally.  Such ensembles clearly have a rich phenomenology
and could potentially give rise to characteristic signals in the kinematic distributions
of event-shape variables at colliders and as well as in distinctive recoil-energy spectra observed at
direct-detection experiments.

Finally, near the end of Sect.~\ref{sec:RandomMatrixGeneration},
we stated that a single universe can only exhibit one mass spectrum for the dark sector.
In other words, as stated there, we get only ``one roll of the dice''.
Of course, there are certain situations in which this is not strictly true.  
For example, if the random effects which ultimately give rise to our dark-sector mass spectrum 
are associated with a first-order symmetry-breaking phase transition,
then different regions of spacetime could each correspond to their own distinct roll of the dice.
Likewise, multiple rolls of the dice could also arise if we live a full-fledged multiverse,
as recent developments in string theory suggest. 
These questions are especially critical in the case of differing dark sectors
because the dark sector carries a significant matter-based energy density and thus its properties play 
a major role in the time evolution of the universe.

In either case, however, the end result would be
a series of different universes (or regions of spacetime) with different dark sectors,
each existing within its own spatial domain and separated from the others 
by domain walls.   Such domain-wall topological defects would undoubtedly give rise to dramatic effects of their
own, even beyond those associated with the properties of the different dark sectors they separate,
and consequently there are tight phenomenological and astrophysical/cosmological
constraints that can already be placed on such scenarios coming from CMB anistropy limits
and other energy-density constraints.
However, if the only differences between the different universes are those that 
result from different random throws of the 
$SU(N)$-breaking dice in the dark sector, then the $n_{\rm SU(N)}$ probability distribution functions
we have discussed in this paper might have an even greater relevance 
than we have imagined here, 
as averages across the entire universe or multiverse as a whole.


\begin{acknowledgments}


We would like to thank Jonathan Feng and Shrihari Gopalakrishna for useful discussions.
KRD is supported in part by the Department of Energy under Grant DE-FG02-13ER41976
and by the National Science Foundation
through its employee IR/D program.  JK is supported in part by NSF CAREER grant 
PHY-1250573.  BT is supported in part by an internal research award from Reed College.   
KRD, JK, and BT would also like to thank the Center for Theoretical Underground Physics 
and Related Areas (CETUP$^\ast$) in Lead, South Dakota, for hospitality during the 2015 Summer Program.  The opinions
and conclusions expressed herein are those of the authors, and do not represent any 
funding agencies.

\end{acknowledgments}


\end{document}